\documentclass[aps,pra,reprint,showpacs]{revtex4-1}

\usepackage{watermark}
\usepackage{amsmath}
\usepackage{amsfonts}
\usepackage{graphicx}
\usepackage{epsfig}
\usepackage{epstopdf}

\usepackage{hyperref}
 \hypersetup{
    bookmarks=true,         % show bookmarks bar?
    unicode=false,          % non-Latin characters in AcrobatÕs bookmarks
    pdftoolbar=true,        % show AcrobatÕs toolbar?
    pdfmenubar=true,        % show AcrobatÕs menu?
    pdffitwindow=false,     % window fit to page when opened
    pdfstartview={FitH},    % fits the width of the page to the window
    pdftitle={Positron scattering and annihilation on noble gas atoms},    % title
    pdfauthor={J. A. Ludlow},     % author
    pdfsubject={Positron_noble},   % subject of the document
    pdfcreator={J. A. Ludlow},   % creator of the document
    pdfproducer={Producer}, % producer of the document
    pdfkeywords={keywords}, % list of keywords
    pdfnewwindow=true,      % links in new window
    colorlinks=true,       % false: boxed links; true: colored links
    linkcolor=blue,          % color of internal links
    citecolor=blue,        % color of links to bibliography
    filecolor=magenta,      % color of file links
    urlcolor=blue           % color of external links
}

\usepackage{amsmath}
\usepackage{amsfonts}
\usepackage{graphicx}
\usepackage{epsf}
\usepackage{subfigure}
\newcommand{\etal}{\emph{et al.}~}
\newcommand{\eqn}[1]{Eq.~(\ref{#1})}
\newcommand{\fig}[1]{Fig.~\ref{#1}}

\newcommand{\eps}{\varepsilon}
\newcommand{\Zeff}{Z_{\rm eff}}
\newcommand{\tj}[6]{\left({#1\atop#4}{#2\atop#5}{#3\atop#6}\right)}
\newcommand{\sj}[6]{\left\{{#1\atop#4}{#2\atop#5}{#3\atop#6}\right\}}

\begin{document}

\title{Positron scattering and annihilation on noble gas atoms}

\author{D.~G.~Green}
\email[Address correspondence to~]{dermot.green@balliol.oxon.org}
\altaffiliation{\newline Present address: Joint Quantum Centre (JQC) Durham-Newcastle, Department of Chemistry, Durham University, South Road, Durham, DH1 3LE, United Kingdom.}
\author{J.~A.~Ludlow}
%\email{john@aquaq.co.uk}
\altaffiliation{Present address: AquaQ Analytics,
Suite 5, Sturgeon Building, 9-15 Queen Street, Belfast, BT1 6EA, UK.} 
\author{G.~F.~Gribakin}
\email{g.gribakin@qub.ac.uk}
\affiliation{Department of Applied Mathematics and Theoretical Physics, Queen's University Belfast, Belfast BT7 1NN, Northern Ireland, United Kingdom}

\begin{abstract}
Positron scattering and annihilation on noble gas atoms is studied \emph{ab initio} using many-body theory methods for positron energies below the positronium formation threshold. We show that in this energy range the many-body theory yields accurate numerical results and provides a near-complete understanding of the positron-noble-gas-atom system. It accounts for positron-atom and electron-positron correlations, including the polarization of the atom by the positron and the non-perturbative process of virtual positronium formation. These correlations have a large effect on the scattering dynamics and result in a strong enhancement of the annihilation rates compared to the independent-particle mean-field description.
Computed elastic scattering cross sections are found to be in good agreement with recent experimental results and Kohn variational and convergent close-coupling calculations. 
The calculated values of the annihilation rate parameter $\Zeff$
(effective number of electrons participating in annihilation) rise steeply along the sequence of noble gas atoms due to the increasing strength of the correlation effects, and agree well with experimental data. 

\end{abstract}

\pacs{34.80.Uv, 78.70.Bj}

\maketitle

%\textsc{TO DO:  DGG to re-write analytic expressions in appendix; discuss contribution/diminishing values of higher order diagrams that we have not considered}

%\newpage

%******************************************************************

\section{Introduction}
The scattering of low-energy positrons from noble gas atoms has been
the subject of theoretical studies for many decades~\cite{massey}. For an overview 
of the field of low-energy positron scattering, see the review~\cite{gribakinrev}. Although the exchange interaction is
absent, positron scattering from atoms is considerably more challenging to treat theoretically than the related 
problem of electron-atom scattering. 
For positrons, the static interaction with the atom is repulsive. 
At low incident positron energies the attractive polarization potential induced by the positron on the atom overcomes the static repulsion, leading to a delicate balance between the opposing potentials. 
A key role is thus played by positron-atom and positron-electron correlations.
In addition, phenomena unique to positrons occur, namely, positronium formation (virtual or real) and positron annihilation. 

Positronium (Ps) formation is a process in which a positron captures an atomic electron into a bound state. It occurs when the positron energy exceeds the Ps-formation threshold $\eps_{\rm Ps}=I+E_{1s}({\rm Ps})=I-6.8$~eV, where $I$ is the ionization potential of the atom and $E_{1s}({\rm Ps})$ is the ground-state energy of Ps. In positron-atom collisions this is usually the first inelastic channel to open, and it has a pronounced effect on positron scattering~\cite{gribakinrev}. Ps formation also affects the positron-atom interaction at energies below $\eps_{\rm Ps}$, where it is \textit{virtual}.
Besides elastic scattering, another channel open at all positron energies is
positron annihilation. For atomic and molecular targets the positron annihilation cross section is traditionally parameterized as $\sigma _a=\pi r_0^2(c/v)\Zeff$, where $r_0$ is the classical electron radius, $c$ is the speed of light, $v$ is the incident positron velocity, and $\Zeff$ is the effective number of electrons participating in the annihilation process. For $\Zeff=1$ this formula gives the basic electron-positron annihilation cross section in the nonrelativistic Born approximation \cite{QED}. For many-electron targets $\Zeff$ may naively be expected to be close to the number of electrons in the atom.
However, the positron-atom interaction and electron-positron correlations have a strong effect on the annihilation rates \cite{Pomer,GS64}. Experimental studies of positron annihilation in heavier noble gases yield $\Zeff$ values that are orders of magnitude greater than those obtained in a simple static-field approximation~\cite{surko,coleman75}.

In this paper we use diagrammatic many-body theory to describe the interaction of positrons with noble gas atoms. Many-body theory allows one to understand and quantify the role and magnitude of various correlation effects.
Scattering phase shifts, differential and total elastic scattering cross sections, and $\Zeff$ are calculated \emph{ab initio} with proper inclusion of the correlations \cite{ludlowphd}. 
Excellent agreement with experimental results and the results of other sophisticated theoretical approaches is found. This work, taken together with the many-body theory calculations of $\gamma$-spectra and rates for annihilation on core electrons of noble gases~\cite{DGGinnershells}, forms a comprehensive study that provides a near-complete understanding of the positron-noble-gas-atom system at positron energies below the Ps-formation threshold.

Many-body theory~\cite{fetter} provides a natural framework for the inclusion of electron-electron and electron-positron correlations. It uses the apparatus of quantum field theory to develop a perturbative expansion for the amplitudes of various processes. The ability to show various contributions pictorially
by means of Feynman diagrams makes the theory particularly transparent and helps one's intuition and understanding of many-body quantum phenomena.
This theory is ideally suited to the problem of a particle interacting with a closed-shell atom, with successful applications to electron scattering from noble-gas atoms (see, e.g., Refs.~\cite{kelly,amusiaelect,Amusia2,wrjon,boyle}). The study of positron-atom scattering using the many-body theory thus should have been straightforward. 
However, progress in this direction was stymied by the difficulty in accounting for virtual Ps formation, as the Ps bound state cannot be accurately described by a finite number of pertubation terms. The need for
this was realised early on \cite{rpa1}, but a proper solution including summation of an infinite series of ``ladder'' diagrams was achieved only much later \cite{hydrogen}. The effect of virtual Ps formation nearly doubles the strength of the positron-atom correlation potential, as the terms in the series are of the same sign, leading to a large total. In contrast, in electron-atom scattering, such series is sign-alternating, giving a small, often negligible, overall contribution.

The first application of the many-body theory to positron scattering was for helium \cite{rpa1}. This study accounted for polarization of the target by the positron and demonstrated the importance of
virtual Ps formation by using a rather crude approximation  (see also \cite{rpa1_2002}). This approximation was also used in subsequent studies for helium and other noble-gas atoms \cite{rpa2,rpa3}. A more sophisticated approximation to the virtual Ps contribution was 
developed and applied to positron scattering, binding and annihilation~in Refs.~\cite{unsw1,unsw4,unsw2,unsw3} (see also \cite{Dzuba}). It was later used to calculate real Ps formation in noble-gas atoms \cite{DG06} and produced mixed results. The complete evaluation of the ladder-diagram series was implemented in the positron-hydrogen study \cite{hydrogen} which used B-spline basis sets to discretize the positron and electron continua. 
This approach has since been applied to positron binding to the halogen negative ions~\cite{halogen}, to positron scattering and annihilation on hydrogen-like ions~\cite{DGGhlike}, to the calculation of gamma-ray spectra for positron annihilation on the core and valence electrons in atoms~\cite{spectra,DGGinnershells} and in molecules~\cite{DGGmoleculargamma}.
Another many-body theory technique that allows one to sum the dominant series of diagrams to all orders is the linearized coupled-cluster method
which was used to calculate positron-atom bound states for a large number of atoms \cite{DFGH12,HDF14}.

Recently, a series of high-quality experimental measurements and convergent close-coupling (CCC) calculations have been performed for low-energy positron scattering along the noble gas sequence \cite{sullivanhe,zecca,zecca1,zeccaxe,sullivannear,sullivankr,sullivanxe,bray2012}. 
In the light of these new data, the many-body theory approach developed by the authors is applied here to a thorough study of positron interaction with the noble gas atoms.  

The rest of the paper is organized as follows. In Secs. \ref{sec:theory} and \ref{sec:implem} we describe the many-body theory and the numerical implementation of this theory. In Sec. \ref{sec:results} we present 
results for the scattering phase shifts, differential and total elastic scattering cross sections, and the annihilation parameter $\Zeff$ (both energy-resolved and thermally averaged), and compare with existing experimental and theoretical data.
We conclude with a brief summary and outlook. 
Algebraic expressions for the many-body diagrams are provided in Appendix
\ref{app:diag} and tabulated numerical results are in Appendix \ref{app:res}. 
We use atomic units (a.u.) throughout, unless stated otherwise. 

%**********************************************************************
\section{Many-body theory}\label{sec:theory}

\subsection{Dyson equation and self-energy}
The many-body-theory description of a positron interacting with an atom is based on the Dyson equation (see, e.g., Ref. \cite{Mig}),
\begin{equation}\label{Dyson44}
(H_0 + \Sigma_{\eps})\psi_{\eps}=\eps\psi_{\eps} ,
\end{equation} 
where $\psi_{\eps}$ is the (quasiparticle) wave function of the positron, $H_0$ is the zeroth-order Hamiltonian of the positron in the static field of the atom (usually described in the Hartree-Fock (HF) approximation), $\eps$ is the positron energy, 
and $\Sigma_{\eps}$ is a nonlocal, energy-dependent correlation potential. This potential is equal to the self-energy of the single-particle Green's function of the positron in the presence of the 
atom \cite{Bell}. It incorporates the many-body dynamics of the positron-atom interaction. As the potential $\Sigma_{\eps}$ is nonlocal, Dyson's equation is an integral equation,
\begin{equation}\label{Dyson41}
H_0\psi _\eps ({\bf r})+ \int \Sigma_{\eps}({\bf r}, {\bf r}')\psi_{\eps}  
({\bf r}') d{\bf r}'=\eps \psi_{\eps}({\bf r}).
\end{equation}
The correlation potential $\Sigma_{\eps}$ can be evaluated as an infinite perturbation series in powers of the 
residual electron-positron and electron-electron interactions. Because of the spherical symmetry of the atomic potential, \eqn{Dyson41} can be solved separately for each partial wave of the incident positron. 

\begin{figure}[t!!]
\begin{center}
\includegraphics*[width=0.48\textwidth]{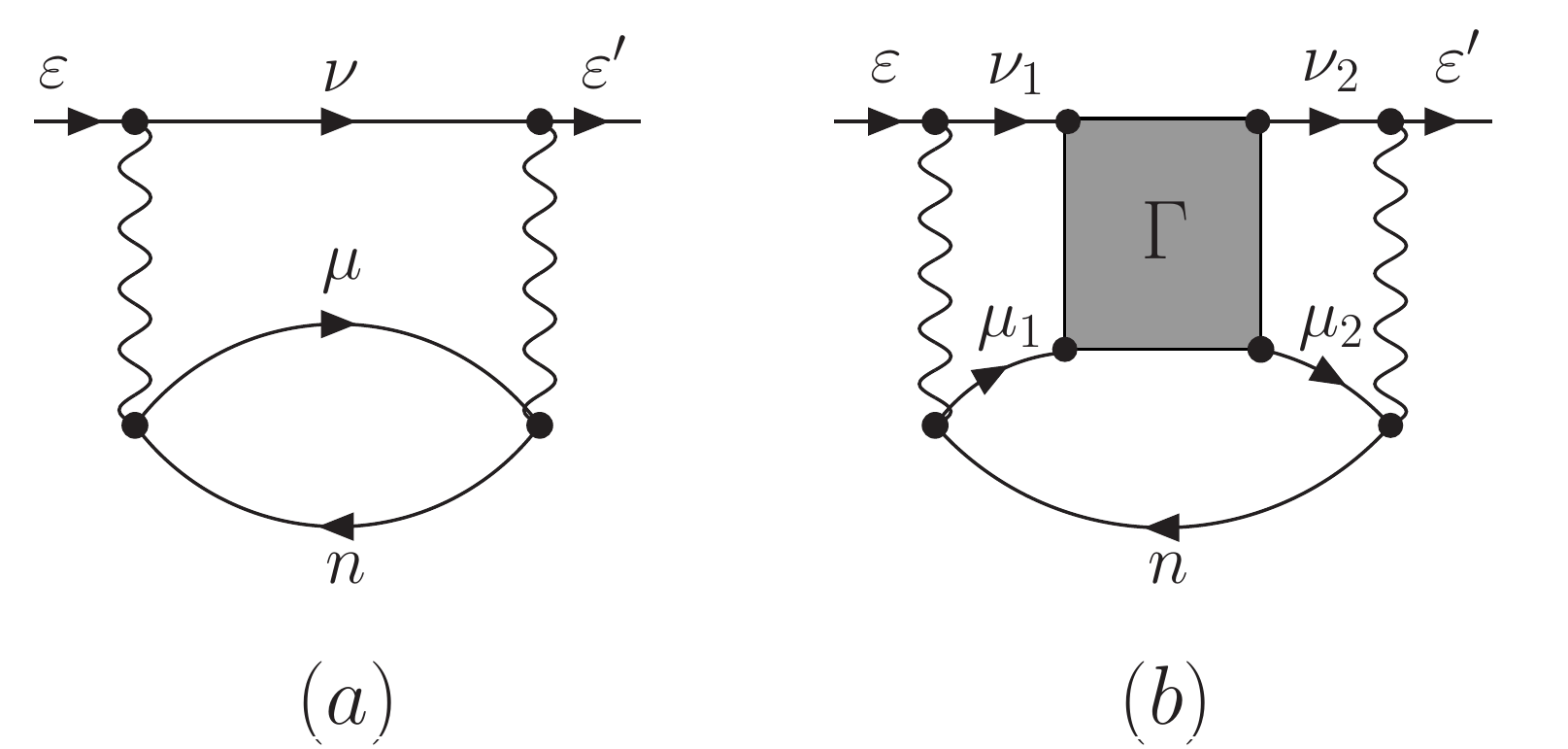}
\end{center}
\caption{Main contributions to the positron self-energy $\Sigma _\eps $. 
The lowest, second-order diagram (a), $\Sigma ^{(2)}_\eps$, describes the effect of polarization; diagram (b), $\Sigma ^{(\Gamma )}_\eps $, accounts
for virtual Ps formation represented by the $\Gamma $-block. Top lines in
the diagrams describe the incident and intermediate-state positron. Other lines with the arrows to the right
are excited electron states, and to the left, holes, i.e., electron states
occupied in the target ground state. 
Wavy lines represent Coulomb interactions.
Summation over all intermediate  positron, electron and hole states is assumed.\label{fig:Sig_2Ps}}
\end{figure}

The main contribution to the positron self-energy $\Sigma_{\eps}$ is given
by the two diagrams shown in \fig{fig:Sig_2Ps}. The second of these in fact represents an infinite subsequence of diagrams which describe virtual Ps formation. For a positron interacting with a one-electron target (the hydrogen atom or hydrogen-like ion), the diagrams shown in \fig{fig:Sig_2Ps} constitute a complete expansion of $\Sigma _\eps$. The algebraic expressions for these diagrams can be found in \cite{hydrogen}.

Diagram \ref{fig:Sig_2Ps}\,(a), $\Sigma^{(2)}_\eps $, accounts for the polarization of the atom by the positron. At large positron-atom separations this diagram has the asymptotic behaviour,
\begin{equation}
\Sigma^{(2)}_{\eps}({\bf r},{\bf r}')\approx
-\frac{\alpha _d e^2}{2r^4}\delta({\bf r}-{\bf r}') ,
\end{equation}
where $\alpha _d$ is the static dipole polarizability of the atom (here, in the HF approximation). This second-order diagram (with its exchange counterparts) is known to provide a good approximation for $\Sigma_{\eps}$ in electron-atom scattering (e.g., for argon \cite{Amusia2}
or xenon \cite{wrjon}). However, the same approximation in positron-atom scattering is seriously deficient \cite{rpa1,Dzuba}.

\begin{figure*}[t!]
\begin{center}
\includegraphics*[width=0.98\textwidth]{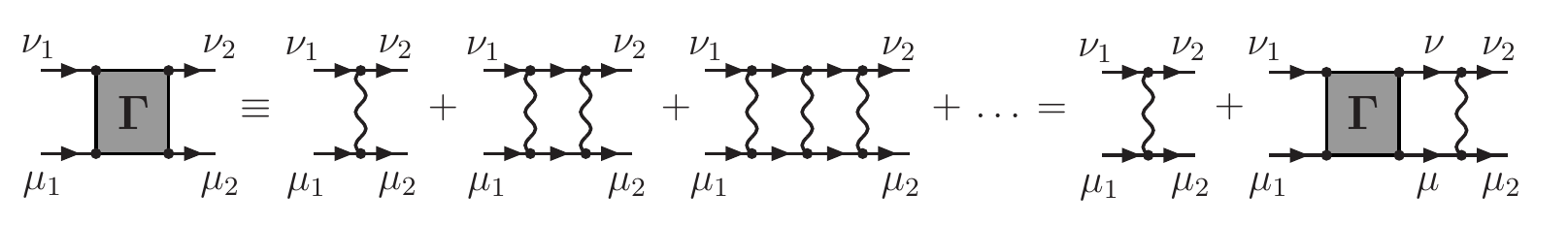}
\end{center}
\caption{Electron-positron ladder diagram series and its sum, the vertex
function $\Gamma $ (shaded block). Comparison between the left- and right-hand
sides of the diagrammatic equation shows that $\Gamma $ can be found by solving a linear equation, \eqn{ladder1}.}
\label{fig:lad}
\end{figure*}

Diagram \ref{fig:Sig_2Ps}\,(b), which we denote $\Sigma^{(\Gamma )}_\eps $, describes the short-range attraction between the positron and the atom due
to virtual Ps formation. The shaded block $\Gamma $ represents the 
sum of electron-positron ladder diagrams, referred to as the vertex function. It satisfies a linear integral equation represented diagrammatically in \fig{fig:lad} and written in the operator form as
\begin{equation}\label{ladder1}
\Gamma = V + V\chi\Gamma ,
\end{equation} 
where $\Gamma$ is the vertex function, $V$ is the electron-positron
Coulomb interaction and $\chi$ is the propagator of the intermediate electron-positron state.
With the electron and positron continua discretized as described in Sec.~\ref{sec:implem}, $\Gamma $ and $V$ become matrices, with $\chi $ being a diagonal matrix of energy denominators. In this case \eqn{ladder1} is a linear matrix equation, which is easily solved numerically \cite{hydrogen}.
Such discretization of the electron and positron continua is valid for energies below the Ps formation threshold, for which the electron-positron pair cannot escape to infinity. 

In order to describe the polarization of multi-electron atoms more accurately, a set of third-order diagrams is also included in the calculation of $\Sigma _\eps$. These diagrams, denoted collectively $\Sigma ^{(3)} _\eps$, are shown in \fig{fig:Sig_3}. Algrebraic expressions for these diagrams are given in Appendix~\ref{app:diag}. Diagrams \ref{fig:Sig_3}~(a), (b), (c) and (d) represent corrections to the second-order
polarization diagram \fig{fig:Sig_2Ps}\,(a) due to electron correlations of the type described by the random-phase approximation with exchange \cite{amusia75}. They account for the
electron-hole interaction and screening of the positron and electron Coulomb field. Diagram \ref{fig:Sig_3}~(e) describes the positron-hole repulsion. 

\begin{figure*}[ht!!]
\begin{center}
\includegraphics*[width=0.75\textwidth]{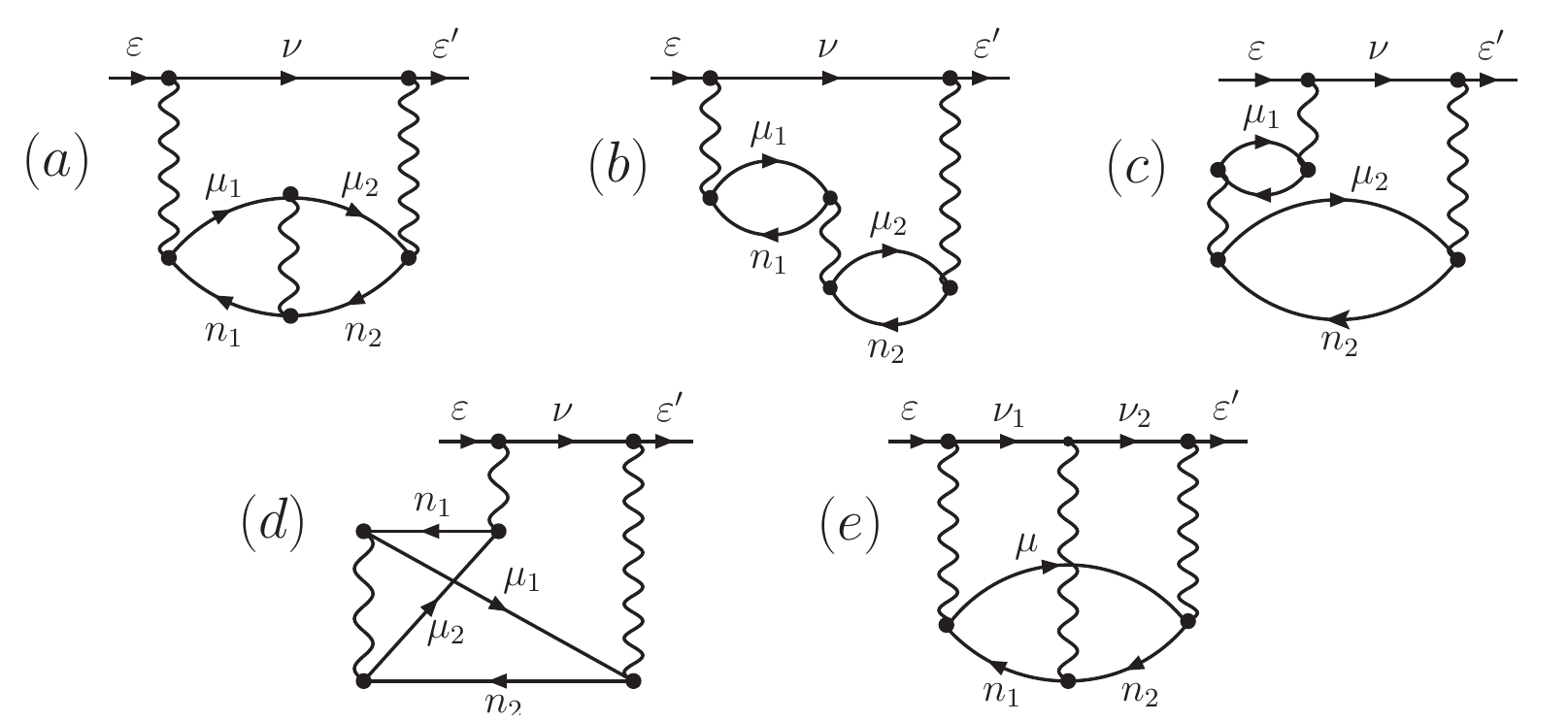}
\end{center}
\caption{Third-order correction diagrams, $\Sigma ^{(3)}_\eps $. Mirror images of the diagrams (c) and (d) are also included. In all diagrams the top, horizontal lines represent the positron.}
\label{fig:Sig_3}
\end{figure*}

Instead of computing the self-energy $\Sigma_{\eps}({\bf r}, {\bf r}')$ in the coordinate representation, it is more convenient to deal with its
matrix elements
\begin{equation}\label{irrud}
\langle\eps'|\Sigma_E|\eps\rangle=\int\varphi_{\eps'}^{*}({\bf r}')
\Sigma_E({\bf r}',{\bf r})\varphi_{\eps}({\bf r})d{\bf r}d{\bf r}',
\end{equation}
with respect to the zeroth-order static-field positron wave functions $\varphi_{\eps}$ with a given orbital angular momentum $\ell$. The latter are eigenstates of the zeroth-order Hamiltonian,
\begin{equation}
H_0\varphi_{\eps}=\eps\varphi_{\eps},
\end{equation}
which satisfy the correct boundary conditions and are appropriately normalized. For true continuous-spectrum positron states the radial wave functions are normalized to a $\delta$-function of energy in Rydberg, $\delta(k^2-k'^2)$. This corresponds to the asymptotic behaviour $P_{\eps \ell}^{(0)}(r)\simeq (\pi k)^{-1/2}
\sin(kr-{\ell\pi}/{2}+\delta^{(0)}_{\ell})$, 
where $\delta^{(0)}_{\ell}$ are the static HF-field phase shifts, and $k$ is the wavenumber related to the positron energy by $\eps =k^2/2$. The intermediate states in the diagrams are square-integrable electron and positron basis functions --
%obtained by diagonalising $H_0$ using 
eigenstates of $H_0$ constructed from
B-splines in a finite-size box of radius $R$ (see Sec.~\ref{subsec:Bspl}).

%**********************************************************************
\subsection{Scattering phase shifts}

The self-energy matrix (\ref{irrud}) can be used to obtain the phase shifts directly \cite{amusia75,phase}. First, a ``reducible'' self-energy matrix $\langle\eps'|\tilde{\Sigma}_E|\eps\rangle$ is found from the integral equation,
\begin{equation}\label{rse}
\langle\eps'|\tilde{\Sigma}_E|\eps\rangle=
\langle\eps'|\Sigma_E|\eps\rangle
+{\cal P}\int
\frac{\langle\eps'|\tilde{\Sigma}_E|\eps^{\prime\prime}\rangle
\langle\eps^{\prime\prime}|\Sigma_E|\eps\rangle}
{E-\eps^{\prime\prime}}d\eps^{\prime\prime},
\end{equation}
where ${\cal P}$ denotes the principal value of the integral. The scattering phase shift is then given by,
\begin{equation}\label{eq:phase}
\delta_{\ell}(k)=\delta^{(0)}_{\ell}(k)+\Delta\delta_{\ell}(k) ,
\end{equation}
where
\begin{equation}\label{eq:tandel}
\tan \left[\Delta\delta_{\ell}(k)\right]=-2\pi
\langle\eps|\tilde{\Sigma}_{\eps}|\eps\rangle ,
\end{equation}
determines the additional phase shift $\Delta\delta_l(k)$ due to positron-atom correlations described by the self-energy.

The reducible self-energy matrix also allows one to find the positron quasiparticle wave function (i.e., solution to the Dyson equation), as
\begin{equation}\label{QP}
\psi_{\eps}({\bf r})=\varphi_{\eps}({\bf r})
+{\cal P}\int \varphi_{\eps'}({\bf r})
\frac{\langle\eps'|\tilde{\Sigma}_{\eps}|\eps\rangle}
{\eps-\eps'}d\eps' .
\end{equation}
Numerically, the integrals in Eqs.~(\ref{rse}) and (\ref{QP}) are calculated using an equispaced positron momentum grid of 200 intervals of $\Delta k=0.02$.
In order for the quasiparticle radial wave function to be correctly
normalized and have the asymptotic behaviour
\begin{equation}\label{nor}
P_{\eps \ell}(r)\simeq\frac{1}{\sqrt{\pi k}}
\sin\left(kr-\frac{\ell\pi}{2}+\delta^{(0)}_{\ell}+\Delta\delta_{\ell}\right) ,
\end{equation}
the wave function obtained from \eqn{QP} must be multiplied by the factor
\begin{equation}\label{nor1}
\cos\Delta\delta_{\ell}=\left[1+
\bigl(2\pi\langle\eps|\tilde{\Sigma}_{\eps}|\eps\rangle\bigr)^2\right]^{-1/2} .
\end{equation}

%**********************************************************************
\subsection{Positron annihilation}

The annihilation rate $\lambda $ for a positron in a gas of atoms or molecules with number density $n$ is usually parameterized by
\begin{equation}\label{Zeff}
\lambda =\pi r_0^2cn\Zeff ,
\end{equation}
where $r_0$ is the classical radius of the electron, $c$ is the speed of
light, and $\Zeff$ is the effective number of electrons that participate in the annihilation process \cite{Pomer,Fraser}.
In general the parameter $\Zeff$ is different from the number of electrons in the target atom, $Z$. In particular, as we shall see, positron-atom and electron-positron correlations can make $\Zeff\gg Z$.

Theoretically, $\Zeff$ is equal to the average positron density at the locations of the target electrons, i.e.,
\begin{equation}\label{Zeff1}
\Zeff=\sum_{i=1}^N\int \delta ({\bf r}-{\bf r}_i)
\left|\Psi _{\bf k}({\bf r}_1,\dots ,{\bf r}_N;{\bf r})\right|^2
d{\bf r}_1\dots d{\bf r}_N
d{\bf r},
\end{equation} 
where $\Psi _{\bf k}({\bf r}_1,\dots ,{\bf r}_N;{\bf r})$ is the total wave function which describes the scattering of the positron with momentum ${\bf k}$ by the $N$-electron target. This wave function is normalized at large positron-atom separations to the positron plane wave incident on the ground-state target with the wave function $\Phi _0$:
\begin{equation}\label{eq:large}
\Psi _{\bf k}({\bf r}_1,\dots ,{\bf r}_N;{\bf r})\simeq
\Phi _0({\bf r}_1,\dots ,{\bf r}_N)e^{i{\bf k}
\cdot {\bf r}}.
\end{equation}

\begin{figure*}[t!]
\begin{center}
\includegraphics*[width=0.7\textwidth]{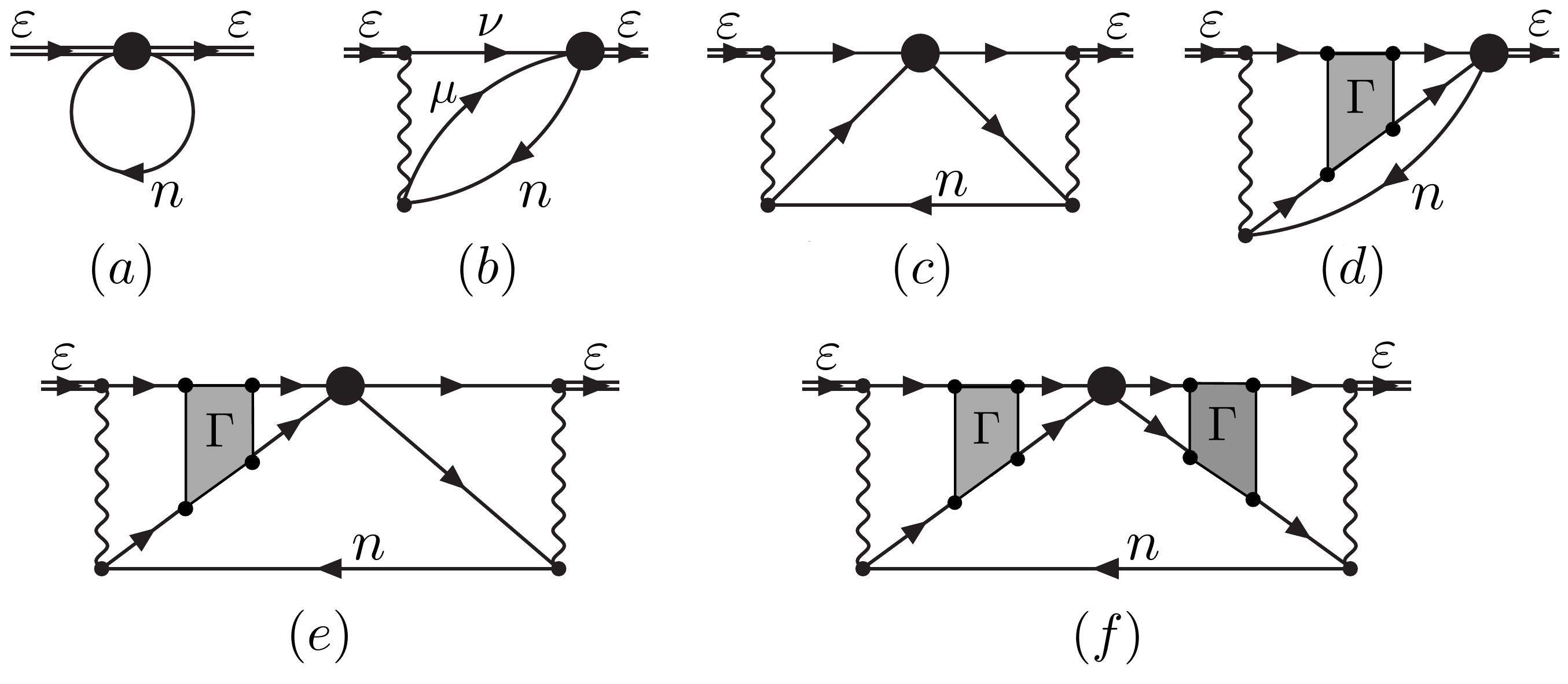}
\end{center}
\caption{Many-body-theory diagrammatic expansion for $\Zeff$.
The solid circle in the diagrams is the $\delta$-function annihilation vertex,
see \eqn{Zeff1}. 
The double lines represent the fully correlated (Dyson) positron quasiparticle wave function of \eqn{QP}, i.e., the HF positron wave function `dressed' with the positron self-energy in the field of the atom.
Diagrams (b), (d) and (e) are multiplied by two to account for their mirror images.}
\label{fig:diagHz}
\end{figure*}

Equation (\ref{Zeff1}) has the form of an amplitude, with the electron-positron delta-function acting as a perturbation. Hence, it is possible to derive a diagrammatic expansion for $\Zeff$ \cite{hydrogen,Dzuba,DG06}.
Figure \ref{fig:diagHz} shows the set of main annihilation diagrams.
In addition to the elements found in the self-energy diagrams, each of the $\Zeff$ diagrams contains one electron-positron $\delta $-function vertex.
The diagrams in \fig{fig:diagHz} provide a complete description of $\Zeff$ for one-electron systems, such as hydrogen and hydrogen-like ions \cite{DGGhlike}. Algebraic expressions for these diagrams can be found in \cite{hydrogen}. The simplest, zeroth-order diagram, \fig{fig:diagHz}\,(a), corresponds to
\begin{equation}\label{eq:Zeff0}
\Zeff^{(0)}=\sum_{n=1}^N\int |\varphi _n({\bf r})|^2|\psi _\eps ({\bf r})|^2d{\bf r},
\end{equation}
i.e., the overlap of the electron and positron densities ($\varphi _n$ being the $n$th electron HF ground-state orbital). It gives $\Zeff$ in the independent-particle approximation.

For the many-electron systems considered here, it is also important to account
for electron screening in the calculation of $\Zeff$. A series of annihilation diagrams with two Coulomb interactions, similar to the self-energy corrections 
in \fig{fig:Sig_3}, are therefore included, see \fig{fig:diagz}. The corresponding algebraic expressions are given in Appendix~\ref{app:diag}.

\begin{figure*}[ht]
\begin{center}
\includegraphics*[width=0.9\textwidth]{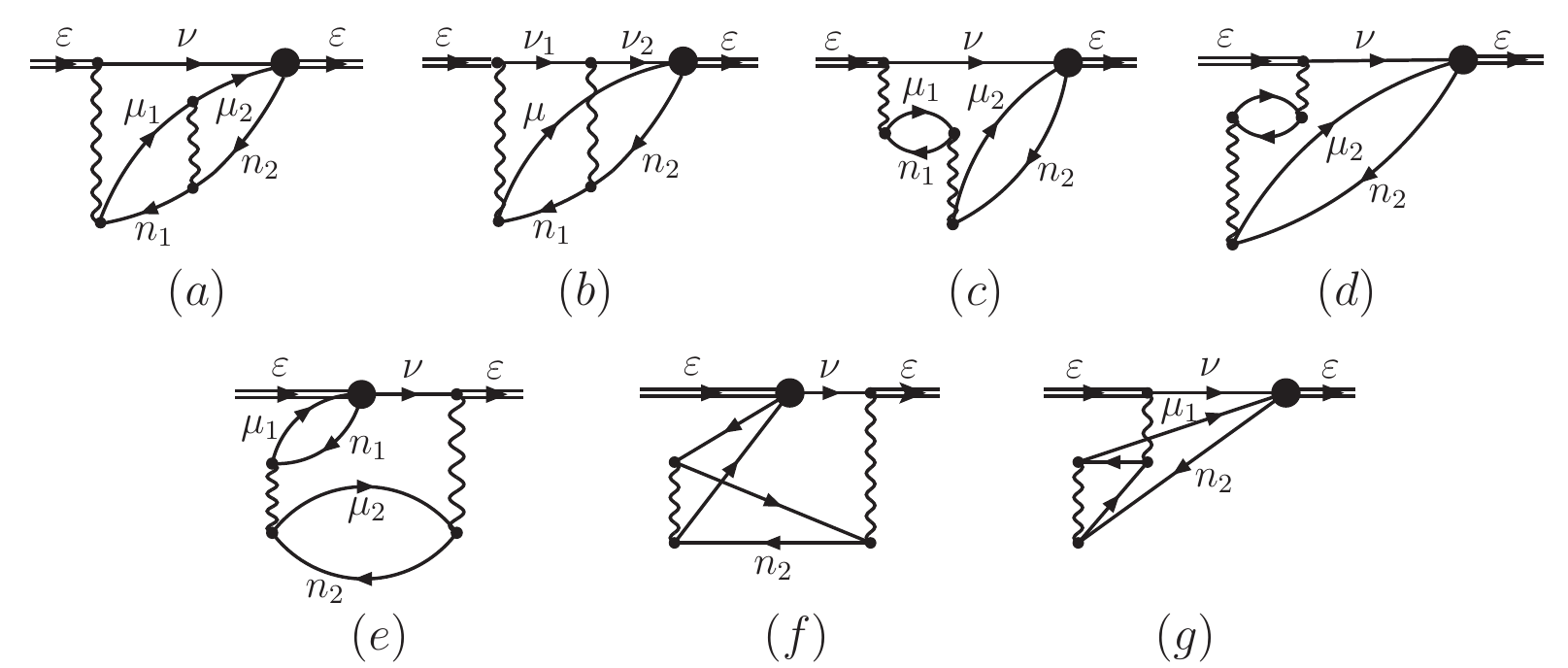}
\end{center}
\caption{Annihilation diagrams which account for corrections to $\Zeff$ due to electron screening of the electron and positron Coulomb field, and electron-hole and positron-hole interactions. The top, horizontal lines represent the positron. The double lines represent the fully correlated (Dyson) positron quasiparticle wave function of \eqn{QP}. All the diagrams have equal mirror images.}
\label{fig:diagz}
\end{figure*}

The external lines in the $\Zeff$ diagrams represent the wave function of the incident positron. In the lowest approximation, one can use the  positron wave function in the static field of the HF ground-state atom, i.e., set $\psi _\eps =\varphi _\eps $, neglecting the effect of the correlation potential $\Sigma _\eps $ on the positron. This effect is in fact quite strong, so to obtain accurate $\Zeff$ one needs to use the quasiparticle positron wave function of \eqn{Dyson44}. Figures \ref{fig:diagHz} and \ref{fig:diagz} represent the latter case, with double lines corresponding to the fully correlated (Dyson) positron quasiparticle wave function obtained from \eqn{QP}, i.e., the HF positron wave functions `dressed' with the positron self-energy in the field of the atom (see Figs.~\ref{fig:Sig_2Ps} and \ref{fig:Sig_3}), and normalized by the factor (\ref{nor1}).

In order to implement the correct normalization of the incident positron wave function to the plane wave $e^{i{\bf k}\cdot {\bf r}}$ at large distances, it is necessary to multiply the
$\Zeff$ diagrams computed for the positron wave function with angular momentum $\ell$, by
\begin{equation}
\frac{4\pi^2}{k}(2\ell+1).
\end{equation}

%\newpage
\section{Numerical implementation}\label{sec:implem}

Below we outline the numerical implementation of the many-body theory
methods described in Sec.~\ref{sec:theory}.

\subsection{B-spline basis}\label{subsec:Bspl}

First, the Hartree-Fock ground state of the atom is calculated with a standard HF package \cite{atomprog}. Using these numerical wave functions, direct and exchange potentials are constructed and the atomic HF Hamiltonian for the positron (i.e., without exchange) or electron (with exchange) is then diagonalized in a B-spline basis \cite{Deboor,Johnson}.
The corresponding eigenvectors are used to construct the positron and electron wave functions. This provides effectively complete sets of positron and electron basis states covering both bound states and the positive-energy continuum \cite{hydrogen,DGGhlike}. These states are then used to calculate the Coulomb and $\delta$-function matrix elements 
(\ref{dir}) and (\ref{dirz}), and to evaluate the many-body diagrams by summing over intermediate electron and positron states.

For the calculations reported here sets of 40 splines of order 6 in a
box of size $R=30$~a.u. were used.  Two outermost subshells are included when calculating the self-energy and annihilation diagrams (Figs. \ref{fig:Sig_2Ps},
\ref{fig:Sig_3}, \ref{fig:diagHz}, and \ref{fig:diagz}). The diagrams are evaluated at 8 energy points from zero incident positron energy up to the Ps-formation threshold and then interpolated onto the required energies. 
The contributions of inner shells to $\Zeff$ are calculated using the diagrams of Fig. \ref{fig:diagHz} \cite{DGGinnershells}.

There is a point concerning boundary conditions satisfied by the B-spline basis states that affects the calculation of the self-energy matrix $\langle\eps'|\Sigma_{E}|\eps\rangle$. The self-energy matrix is evaluated initially using the B-spline basis states $|i\rangle$ as 
$\langle i|\Sigma_{E}|j \rangle$. The number of B-spline basis states used in 
each partial wave ($\sim 15$) is much smaller than the number of continuous spectrum 
states required for an accurate solution of \eqn{rse}. The change to 
$\langle\eps'|\Sigma_{E}|\eps\rangle$ can be made using the effective completeness of the B-spline basis on the interval $[0,R]$,
\begin{equation}\label{complete}
\langle\eps'|\Sigma_{E}|\eps\rangle
=\sum_{ij}\langle\eps'|i\rangle
\langle i|\Sigma_{E}|j \rangle\langle j|\eps\rangle ,
\end{equation}
where $\langle\eps|i\rangle$ is the overlap of the HF state
$\eps$ with the B-spline basis state $|i\rangle$. However, unlike the
B-spline states which satisfy the boundary condition $P_{i\ell}(R)=0$, the
continuous spectrum radial wave function $P_{\eps \ell}(r)$ is finite at the boundary $r=R$. To improve numerical accuracy, a weighting function $f(r)=R-r$ is inserted into \eqn{complete}:
\begin{equation}
\label{complete1}
\langle\eps'|\Sigma_{E}|\eps\rangle=
\sum_{ij}\langle\eps'|f|i\rangle
\langle i|f^{-1}\Sigma_{E}f^{-1}|j \rangle\langle j|f|\eps\rangle ,
\end{equation}
with the ``weighted'' self-energy matrix $\langle i|f^{-1}\Sigma_{E}f^{-1}|j \rangle$, being calculated rather than $\langle i|\Sigma_{E}|j \rangle$.

\subsection{Finite box size}

In general, the finite box size may affect the results at low positron momenta $kR\lesssim 1$. In particular, it limits the range of the polarization potential (represented by $\Sigma _\eps$) to distances not
exceeding $R$. This is countered by adding a correction term to the self energy. At large distances the correlation potential is local, energy independent
and of the form $-\alpha _d/2r^4$, where $\alpha _d$ is the static dipole polarizability. The contribution to the self-energy matrix $\langle \eps '|\Sigma _E|\eps \rangle $ from distances outside the box can then be approximated by
\begin{equation}\label{eq:corr}
\int_R^{\infty}P_{\eps' \ell}(r)\left(-\frac{\alpha _d}{2r^4}\right)P_{\eps\ell}(r)\,dr,
\end{equation}
with the radial wave functions given by their asymptotic form,
\begin{equation}
P_{\eps \ell}(r)=r\sqrt{\frac{k}{\pi}}\left[
j_\ell (kr)\cos \delta_{\ell }^{(0)} - 
n_\ell (kr)\sin \delta_{\ell }^{(0)}\right],
\end{equation}
where $j_{\ell}$ and $n_{\ell}$ are the spherical Bessel and Neumann functions. The correction (\ref{eq:corr}) is added to the self-energy matrix calculated using the many-body theory for $r\leq R$.

\subsection{Angular momentum convergence}

The use of B-spline basis sets provides for a fast convergence of the perturbation-theory sums in the self-energy and $\Zeff$ diagrams with respect to the number of intermediate electron and positron states in a given partial wave with angular momentum $l$. However, the sums in the diagrams are restricted to a finite number of partial waves up to a maximum orbital angular momentum $l_{\rm max}$, and the question of convergence with respect to $l_{\rm max}$ needs to be addressed. One solution successfully tested in Ref.~\cite{hydrogen} is to use extrapolation described by the asymptotic formulae \cite{Ludlow},
\begin{equation}\label{asym3}
\delta _\ell (k)=\delta_\ell ^{[l_{\rm max}]}(k)+\frac{A_\ell (k)}{(l_{\rm max}+1/2)^3} ,
\end{equation}
\begin{equation}\label{asym4}
\Zeff(k)=\Zeff^{[l_{\rm max}]}(k)+\frac{B_\ell (k)}{l_{\rm max}+1/2} ,
\end{equation}
where $\delta_\ell^{[l_{\rm max}]}(k)$ and $\Zeff^{[l_{\rm max}]}(k)$ are
the phase shift and annihilation parameter obtained for a given $l_{\rm max}$,
and $A_\ell(k)$ and $B_\ell(k)$ are constants specific to a particular collision target, positron partial wave $\ell $, and momentum $k$.
These constants and the extrapolated values of the phase shift and $\Zeff$
are determined by fitting $\delta_\ell^{[l_{\rm max}]}(k)$ and $\Zeff^{[l_{\rm max}]}(k)$ over a range of $l_{\rm max}$ to Eqs.~(\ref{asym3}) and (\ref{asym4}), respectively.

\begin{figure}[t!]
\includegraphics*[width=0.47\textwidth]{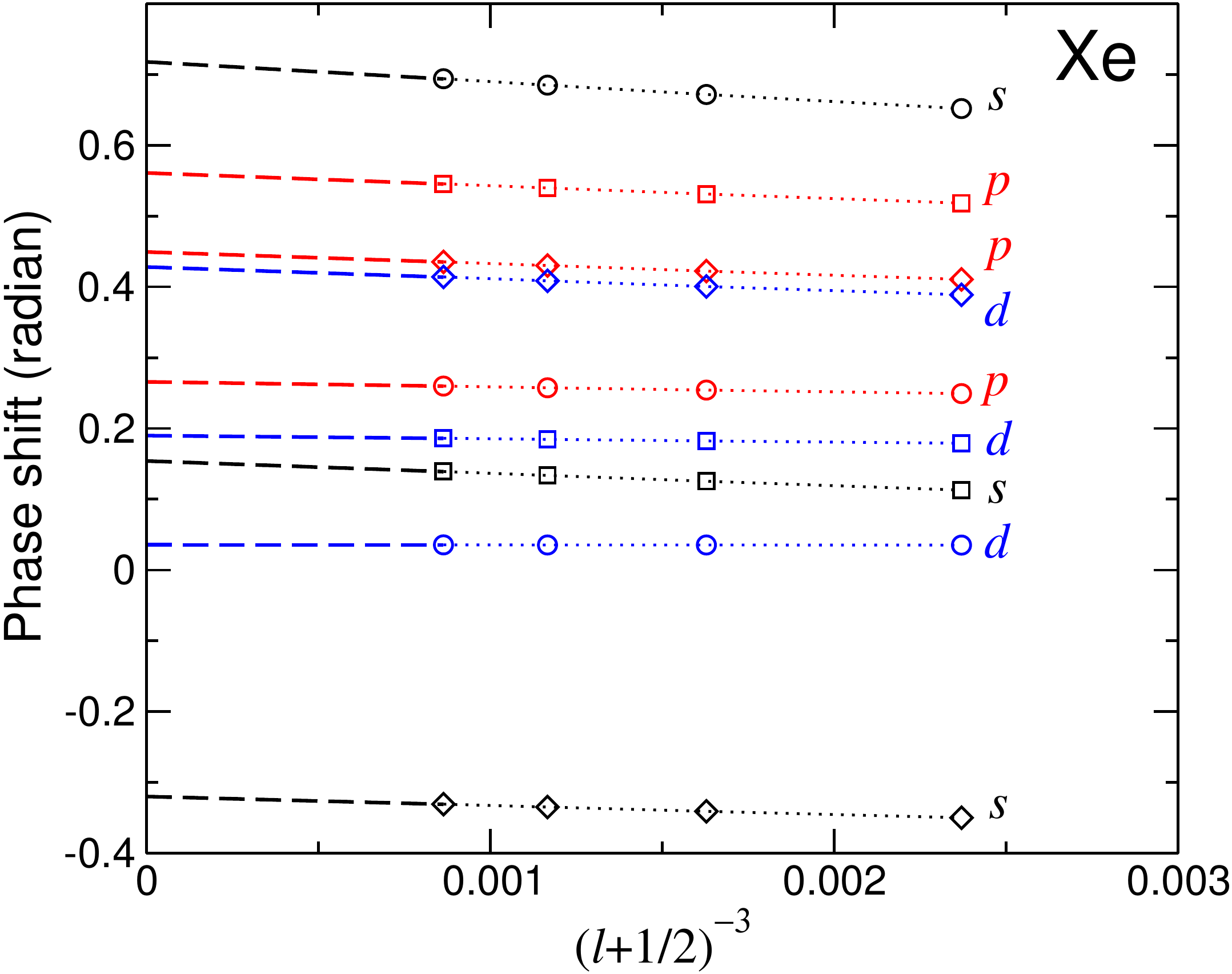}

\includegraphics*[width=0.47\textwidth]{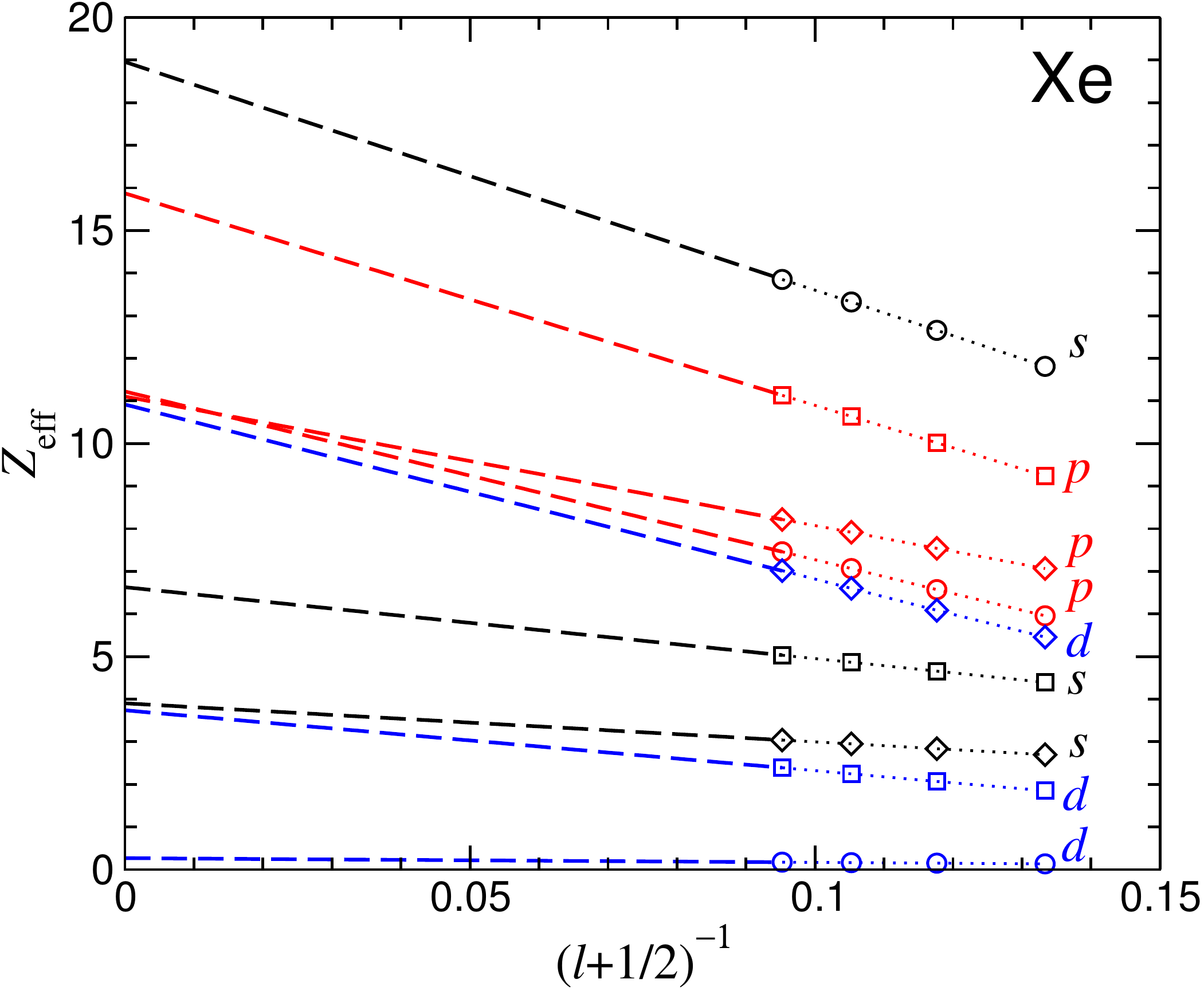}
\vspace*{-5pt}
\caption{Extrapolation of phase shifts and $\Zeff$ for xenon with $s$-, $p$ and $d$-wave incident positrons (black, red and blue 
symbols, respectively) of momenta: $k=0.2$~a.u.~(circles); $k=0.4$~a.u.~(squares); and 
$k=0.6$~a.u.~(diamonds). Symbols show values obtained with $l_{\rm max}=7$--10, dotted lines are shown as a guide only, and dashed lines show extrapolation to $l_{\rm max}\to \infty$.}
\label{xeextrap}
\end{figure}

The use of Eqs.~(\ref{asym3}) and (\ref{asym4}) to extrapolate the phase shifts and $\Zeff$ values to $l_{\rm max}\to \infty$ is illustrated in \fig{xeextrap} for xenon. It shows that the numerical calculations adhere closely to the asymptotic form for $l_{\rm max}=7$--10, allowing a reliable extrapolation to be made. This also indicates that although the extrapolation formulae are derived using perturbation theory \cite{Ludlow}, their use is also valid for the non-perturbative calculations presented here. Note that extrapolation is
particularly important for $\Zeff $, where it contributes up to 30\% of the total. The role of high intermediate orbital angular momenta in $\Zeff$ is large because the annihilation probability is sensitive to small electron-positron separations at the point of coalescence.

\subsection{Partial-wave convergence of elastic scattering cross sections}

The elastic scattering cross section is obtained as a sum over the partial
waves \cite{lan},
\begin{equation}\label{crosssec}
\sigma_{\rm el}=\frac{4\pi}{k^2}\sum_{\ell=0}^{\infty} (2\ell+1)\sin^2\delta_{\ell} .
\end{equation}
At low positron energies only a few partial waves contribute to
$\sigma_{\rm el}$, and the contributions decrease quickly with $\ell$.
On the other hand, the contribution of higher partial waves is more important
in the differential elastic cross section,
\begin{equation}\label{eq:DSC}
\frac{d\sigma_{\rm el}}{d\Omega }=|f(\theta )|^2,
\end{equation}
where
\begin{equation}\label{eq:f}
f(\theta )=\frac{1}{2ik}\sum_{\ell=0}^{\infty} (2\ell+1)(e^{2i\delta _\ell}-1)P_\ell (\cos \theta ),
\end{equation}
is the scattering amplitude and $P_\ell (\cos \theta )$ are Legendre polynomials. Here large $\ell $ interfere constructively at small scattering angles $\theta $ (due to the long-range polarization potential), producing a characteristic cusp at $\theta =0$.

As only $s$-, $p$- and $d$-wave phase shifts have been calculated in the
present work, some way must be found of accounting for the higher partial
waves. This is done by noting that for higher partial waves, the dipole
term in $\Sigma_\eps^{(2)}+\Sigma_\eps^{(3)}$ dominates the self-energy at low energies. At large distances it corresponds to the local energy-independent polarization potential $-\alpha_d/2r^4$. 
It alters the low-energy effective range expansion of the 
scattering phase shifts \cite{omalley},
\begin{align}\label{phase0}
\tan \delta _0&\simeq -ak\left[1-\frac{\pi \alpha _d k}{3a}-\frac{4\alpha _d k^2}{3}
\ln \left( C\frac{\sqrt{\alpha _d}k}{4}\right)\right]^{-1},\\ \label{phasel}
\delta _{\ell}&\simeq \frac{\pi \alpha _dk^2}{(2\ell-1)(2\ell+1)(2\ell+3)},
\quad (\ell\geq 1),
\end{align}
where $a$ is the scattering length and $C$ is a positive constant.
We apply \eqn{phase0} to extract the scattering length from the numerical $s$-wave phase shifts (see Sec.~\ref{subsec:sclen}), and use \eqn{phasel} for $\ell \geq 3$ in the calculations of the differential and total elastic cross sections.

\section{Results: scattering}\label{sec:results} 

\subsection{Phase shifts}

\begin{figure}[t!!]
\includegraphics*[width=0.452\textwidth]{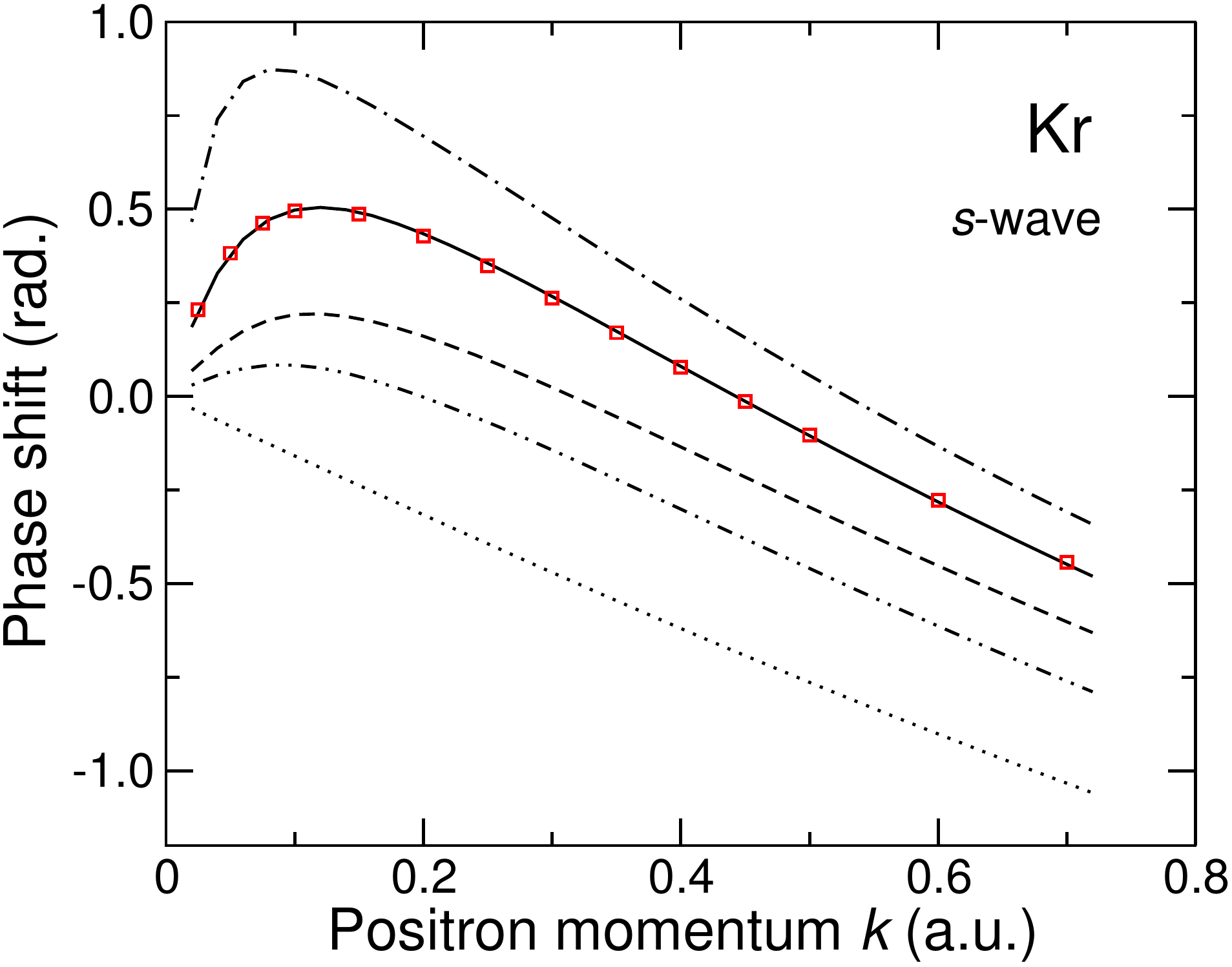}\\[2pt]
\includegraphics*[width=0.452\textwidth]{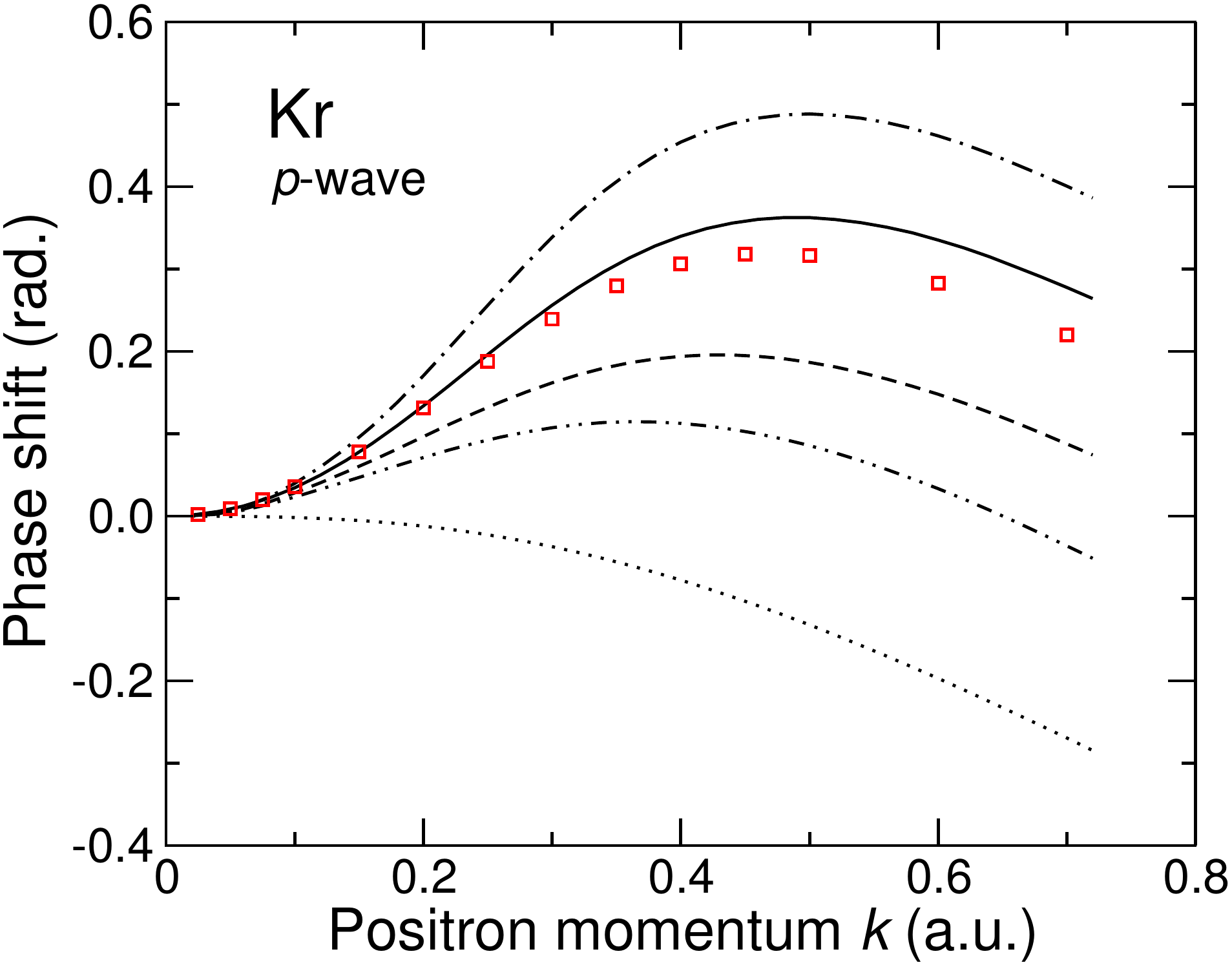}\\[2pt]
\includegraphics*[width=0.452\textwidth]{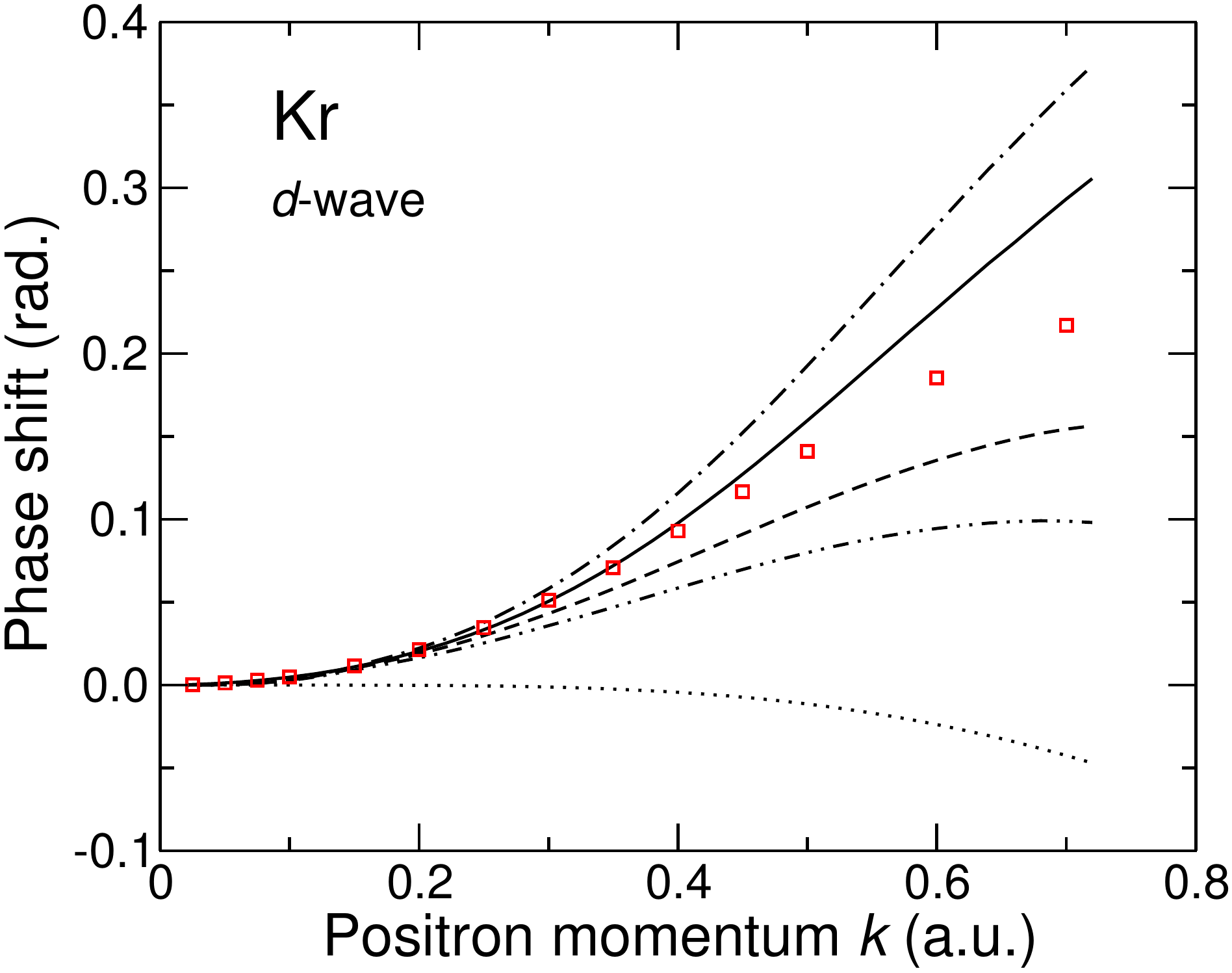}
\caption{Scattering phase shifts for $s$-, $p$- and $d$-wave positrons on Kr in various approximations: HF (static-field) approximation (dotted curve); HF plus second-order correlation 
potential $\Sigma _\eps^{(2)}$ (dashed curve); HF\,+\,$\Sigma _\eps^{(2)}+\Sigma _\eps^{(3)}$ (dot-dot-dash curve); 
HF\,+\,$\Sigma _\eps^{(2)}+\Sigma _\eps^{(\Gamma)}$ (dot-dashed curve); and total: HF\,+\,$\Sigma _\eps^{(2)}+\Sigma _\eps^{(3)}+
\Sigma _\eps^{(\Gamma)}$ (solid curve).
Squares are theoretical results calculated using the polarized orbital method \cite{Mce}.\label{phasekr}}
\end{figure}

Elastic scattering phase shifts for the noble gas atoms are tabulated in appendix \ref{app:res}. 
The general features of the phase shifts as functions of the positron momentum $k$ are illustrated for krypton in \fig{phasekr}. 
In the HF (static-field) approximation, the phase shifts are negative, indicating a repulsive electrostatic field, as is expected for positrons. 
Inclusion of the second-order correlation potential $\Sigma _\eps^{(2)}$ leads to an attractive positron-atom potential at long range, making the phase shifts positive for low $k$ (dashed curves in \fig{phasekr}). 
The asymptotic form of $\Sigma _\eps^{(2)}$ (i.e., $-\alpha_d/2r^4$) leads to terms quadratic in $k$ in the low-energy expansions (\ref{phase0}) and (\ref{phasel}). As a result, the $s$-wave phase shift reaches a maximum and then fall off with increasing $k$, passing through zero (Ramsaur-Townsend effect) to negative phase shifts at higher $k$. The higher-order contributions to the correlation potential ($\Sigma _\eps^{(3)}$ and $\Sigma _\eps^{(\Gamma)}$) have opposing effects. Inclusion of third-order screening diagrams $\Sigma _\eps^{(3)}$ decreases the strength of the positron-atom potential and 
reduces the phase shifts. The contribution of virtual positronium formation $\Sigma _\eps^{(\Gamma)}$ is the greater of the two. It
increases the strength of the positron-atom potential, giving a particulary large contribution at higher $k$ for $p$ and $d$ waves.

In \fig{phasekr} we also show the phase shifts from the polarized-orbital calculations of McEachran {\it et al.}~\cite{Mce} (squares). The polarized-orbital approximation makes a number of drastic assumptions. It considers a linear response of the target to the field of a stationary (i.e., infinitely massive) particle, and drops the monopole polarization term in the potential. Unlike $\Sigma _\eps$, the polarized-orbital potential is local and energy independent, and it does not account for the (nonperturbative) contribution of virtual Ps formation. It is thus remarkable, and likely fortuitous, that the polarized orbital calculations give the $s$-wave phase shift in such close agreement with the many-body calculation. For $p$ and $d$ waves, however, the distinct effect of virtual Ps formation for $k\gtrsim 0.4$~a.u.~produces phase shifts that are 10--20\% greater than those from the
polarized-orbital calculation.

This behaviour of the phase shifts, including the Ramsauer-Townsend minimum in the $s$-wave scattering, is observed for all noble gas atoms. Quantitatively, the correlation effects [i.e., the contribution of $\Delta\delta_{\ell}(k)$ to the phase shift (\ref{eq:phase})] become progressively larger from He to Xe. It was this increase in positron-atom correlational attraction that led to predictions of positron binding to neutral atoms \cite{unsw4}.

\subsection{Scattering lengths}\label{subsec:sclen}

A single quantity that characterizes the strength of positron-atom attraction at low energies is the scattering length $a$. It can be extracted from the effective-range expansion (\ref{phase0}) of the $s$-wave phase shift, written as
\begin{equation}\label{scatl}
\delta _0(k)\simeq -ak -\frac{\pi \alpha _d k^2}{3} - \frac{4a\alpha _d}{3}k^3\ln\frac{\sqrt{\alpha _d}k}{4} + Dk^3,
\end{equation}
where $D$ is a constant. Equation (\ref{scatl}) is convenient for systems in which the scattering length is not too large.
It works well for helium, neon, argon and krypton, and we use it as a two-parameter fit over the range of momenta $k=0.02$--0.06 a.u.~\cite{comm_alpha}.
The corresponding values of the scattering length are shown in Table \ref{stpol}.

\begin{table}[ht!]
\begin{ruledtabular}
\caption{Positron scattering lengths $a$ in a.u.~for the noble gas atoms.}
\label{stpol}
\begin{tabular}{cccccc}
 & He & Ne & Ar & Kr & Xe\\
\hline
$a$\footnote{Present many-body calculations.} &$-0.435$ &$-0.467$ &$-4.41$ &$-9.71$ &$-84.5$\\
$a$\footnote{Previous many-body calculations \cite{unsw2}.} & -- &$-0.43$ &$-3.9$ &$-9.1$ & $\approx -100$\\
$a$\footnote{Polarized orbital calculations \cite{Mce}.}&$-0.53$ &$-0.61$  &$-5.3$ &$-10.4$ &$-45$\\
$a$\footnote{Kohn variational calculations \cite{humscat}.} &$-0.48$ & --& --& --&--\\
$a$\footnote{Convergent close-coupling calculations \cite{bray2012}.} & --& $-0.53$ & $-4.3$ &$-11.2$ & $-117$\\
$a$\footnote{~Experiment: Ar \cite{zecca1}, Kr \cite{zecca}, Xe \cite{zeccaxe}.}& -- & -- & $-4.9\pm 0.7$ & $-10.3\pm 1.5 $ & $-99.2\pm 18.4$
\end{tabular}
\end{ruledtabular}
\end{table}

\begin{figure}[t!!]
\begin{center}
\includegraphics*[clip,width=0.48\textwidth]{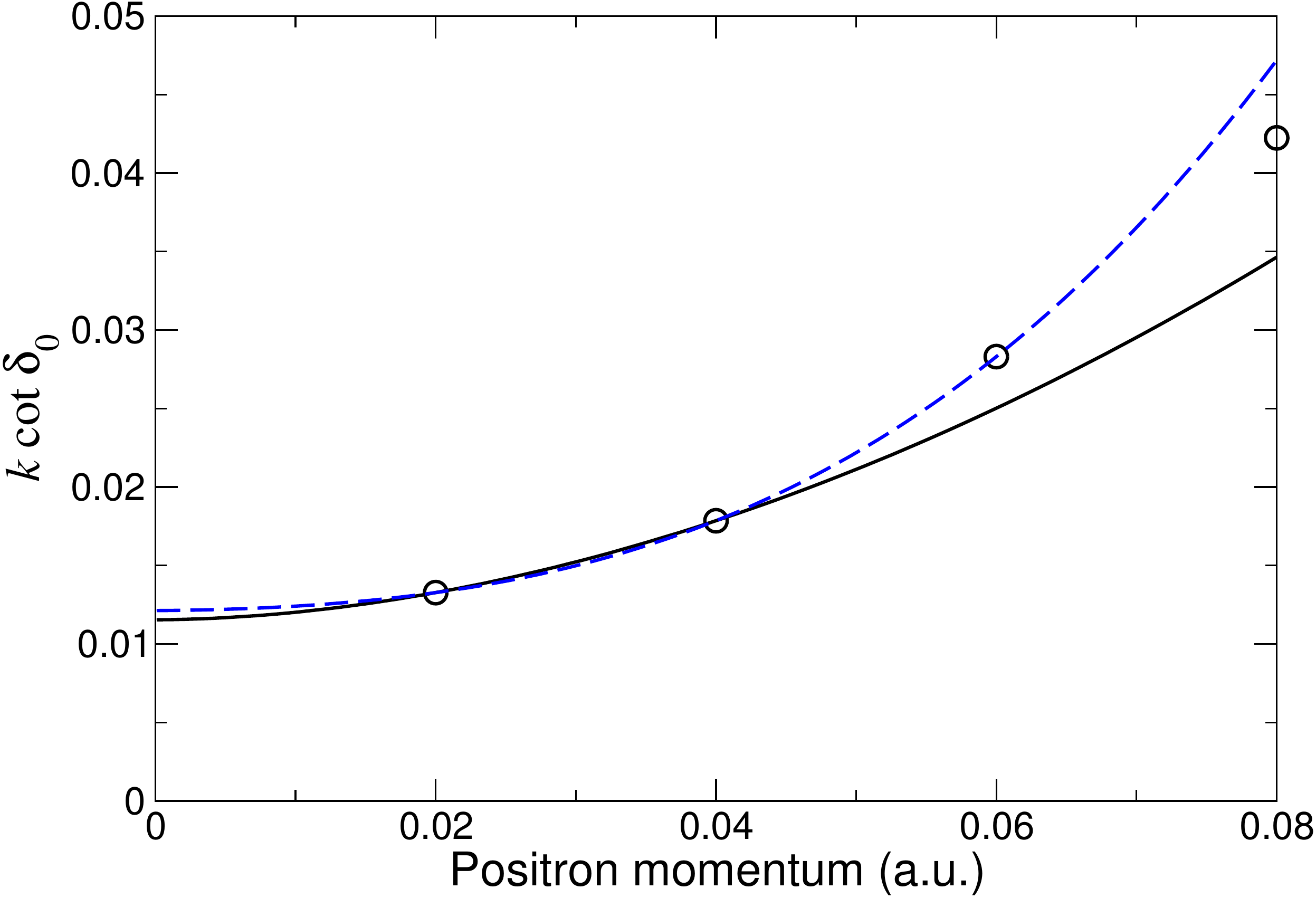}
\end{center}
\caption{$s$-wave scattering phase shift for Xe. Circles show values from the present many-body theory calculation; solid line is a two-parameter fit by \eqn{phase0cot} at $k=0.02$ and 0.04~a.u.; dashed line is a three-parameter fit [\eqn{phase0cot} with a cubic term $Dk^3$ added] using $k=0.02$--0.06~a.u.}
\label{fig:xesphfit}
\end{figure}

Compared with other noble-gas atoms, the $s$-wave phase shift for xenon is
large (see Table \ref{app1e}), indicating a much greater scattering length. In this case it is more convenient to analyse the behaviour of the phase shift using \eqn{phase0} in the form
\begin{equation}\label{phase0cot}
k\cot \delta _0 \simeq -\frac{1}{a}+\frac{\pi \alpha _d k}{3a^2}+\frac{4\alpha _d k^2}{3a}
\ln \left( C\frac{\sqrt{\alpha _d}k}{4}\right).
\end{equation}
Figure \ref{fig:xesphfit} shows the dependence of $k\cot \delta _0$ on the
positron momentum $k$, together with a two-parameter fits using \eqn{phase0cot}, and a three-parameter fit in which a cubic term $Dk^3$ is added on the right-hand-side of \eqn{phase0cot}. The two-parameter fit gives $a=-86.6$~a.u., while the three parameter fit gives $a=-82.4$~a.u. Given the uncertainty of the fitting procedure, our predicted scattering length for Xe is $a=-84.5\pm 2$ a.u. In fact, the exact value of the scattering length for xenon is very sensitive to the positron-atom correlation potential. When the scattering length is large, its reciprocal $\kappa =1/a$ is known to vary linearly with the strength of the potential \cite{lan}. We can thus compare our theoretical prediction $\kappa = -0.0118$~a.u.~with a typical value
$\kappa \sim 0.5$~a.u., compatible with the radius of the Xe atom. (For example, the positron scattering length in the static field of the Xe atom is $a=1.93$~a.u., which corresponds to $\kappa =0.518$~a.u.) This shows that predicting $\kappa $ (and hence, the scattering length) with 1\% accuracy requires a better than 0.1\% accuracy in the calculation of the correlation potential, which is probably beyond any method for such a complex many-electron target as Xe.

If the scattering length $a$ is large and negative then a virtual state exists at the energy $\epsilon =\hbar^2/2ma^2$, where $m$ is the positron mass \cite{lan}. It gives rise to enhanced elastic scattering and annihilation ($\propto |a|^2$) at low positron energies. Table \ref{stpol} compares the scattering lengths extracted from the $s$-wave phase shifts with other theoretical and experimental results. Helium and neon display close scattering lengths in the present calculation, with the result for neon in good agreement with previous many-body theory calculations \cite{unsw2}, although 
somewhat lower than other theoretical predictions \cite{Mce,humscat,bray2012}. 
For argon and krypton, there is close accord between the present many-body theory, polarized orbital \cite{Mce}, and CCC \cite{bray2012} calculations, and the experimental results of Zecca {\it et al.}~\cite{zecca1, zecca}. 
The scattering length increases across the noble-gas atom sequence, giving rise to a virtual $s$-wave level for xenon with the energy of approximately 2~meV. As mentioned above, this indicates that scattering calculations in the low-energy region for xenon will display a high sensitivity to the representation of the correlation potential. The many-body theory scattering length calculated here for xenon is a factor of two greater than that from the polarized-orbital calculations \cite{Mce}, but somewhat lower than those obtained in the earlier many-body calculations \cite{unsw2} and the CCC calculations \cite{bray2012}. However, it is compatible with the experimental value determined by extrapolating the measured low-energy cross section with the aid of CCC calculations \cite{zeccaxe}.

\subsection{Elastic scattering cross sections}

\begin{figure*}[h!p]
\begin{center}
\includegraphics*[clip,width=0.8\textwidth]{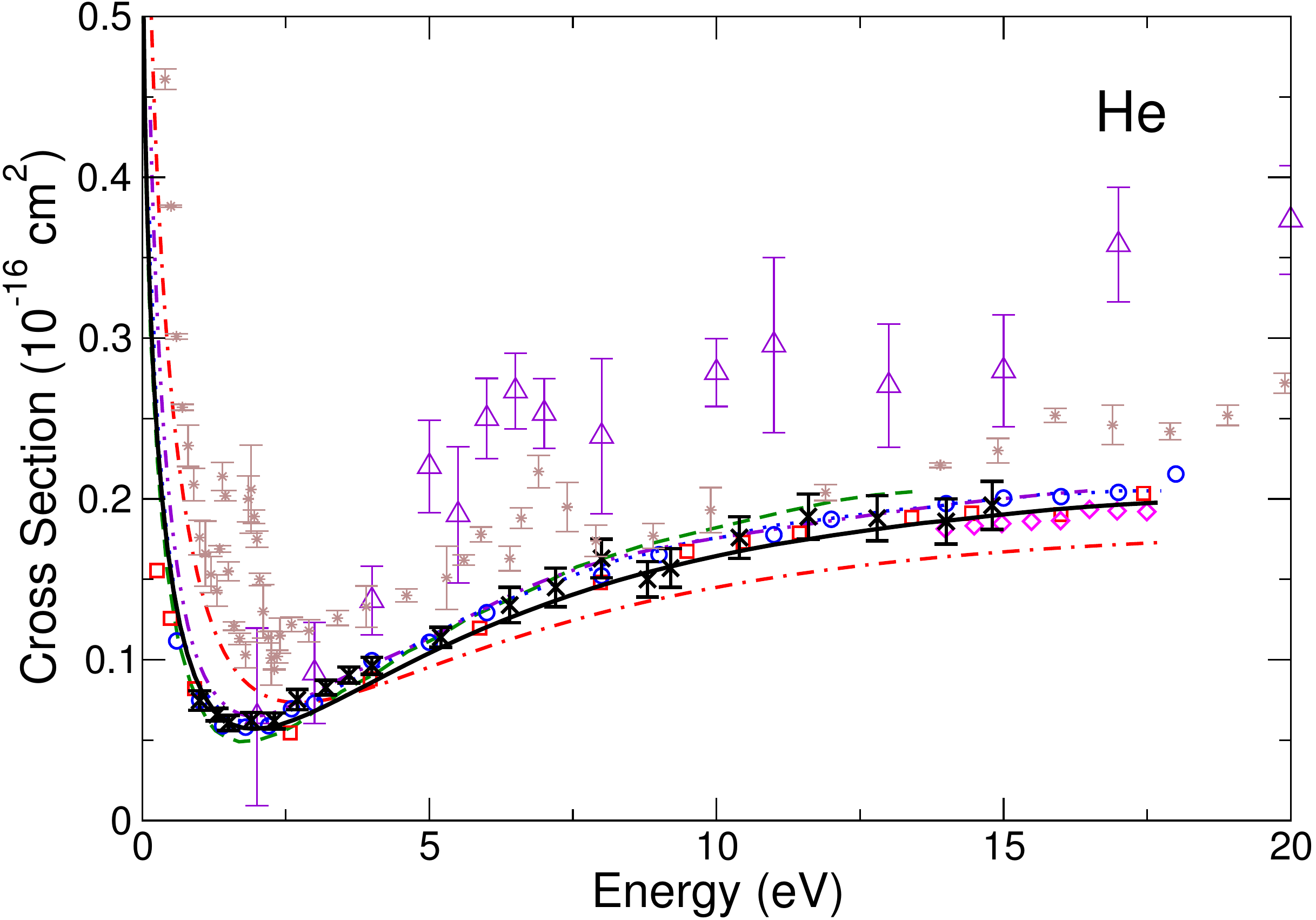}
\end{center}
\caption{Elastic scattering cross section for He.
Theory: solid, present many-body theory; dot-dashed, polarized orbital, \protect\cite{Mce}; 
dashed, MBPT \protect\cite{unsw2}; dotted, CCC \protect\cite{wu}; dot-dot-dashed, 
Kohn variational \protect\cite{humhe}.
Experiment: squares, Stein {\it et al.~}\protect\cite{stein}; circles, 
Mizogawa {\it et al.}~\protect\cite{mizogawa};
stars, Karwasz {\it et al.}~\protect\cite{karwasz_he}; diamonds, Coleman {\it et al.}~\protect\cite{coleman};
crosses, Sullivan {\it et al.}~\protect\cite{sullivanhe}; triangles, Nagumo {\it et al.}~\protect\cite{nagumo_he}.}
\label{fig:crosshe}
%\end{figure*}
%\begin{figure*}[ht!]
\begin{center}
\includegraphics*[clip,width=0.8\textwidth]{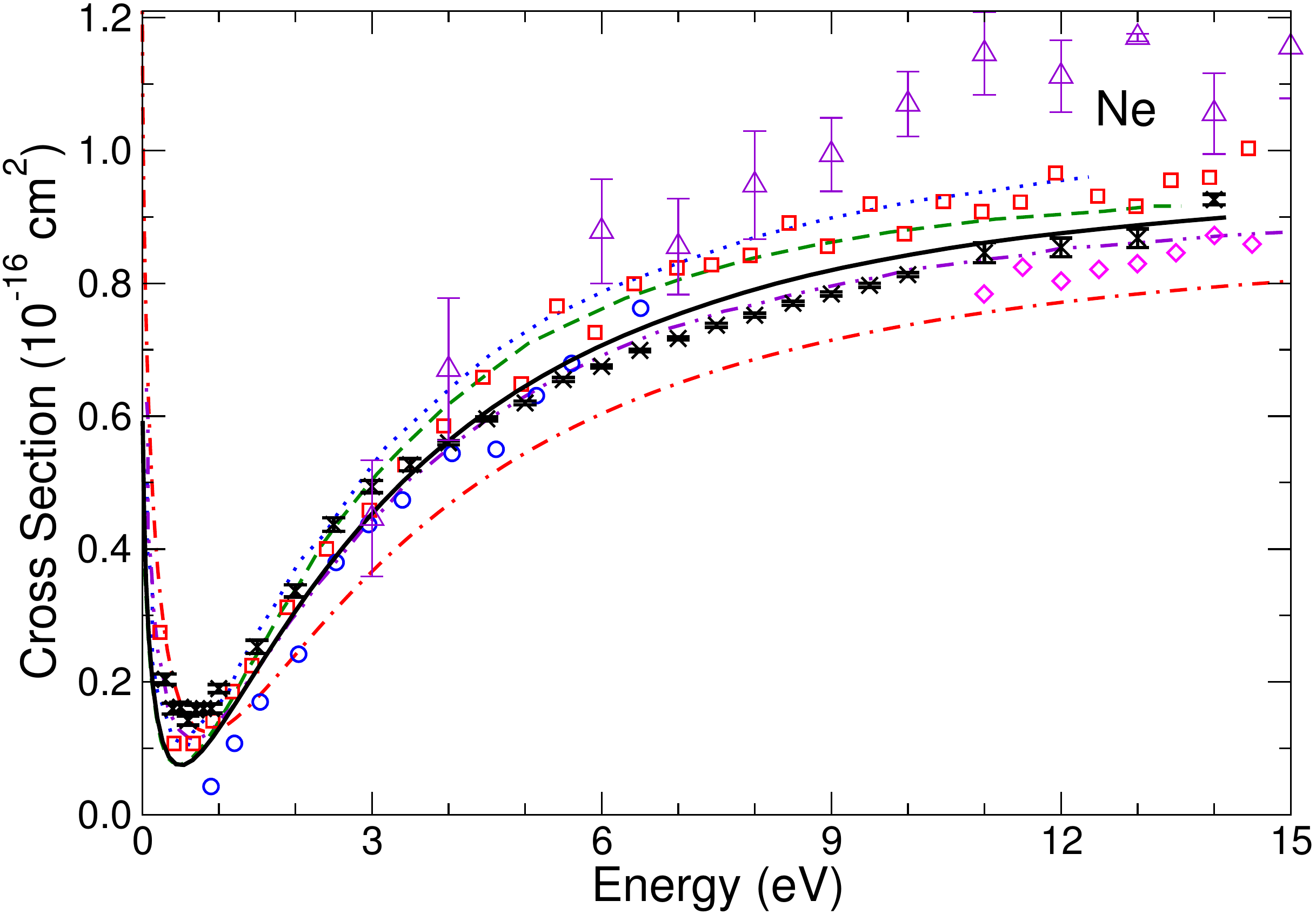}
\end{center}
\caption{Elastic scattering cross section for Ne.
Theory: solid, present many-body theory; dot-dashed, polarized orbital, \protect\cite{Mce}; 
dashed, MBPT \protect\cite{unsw2}; dotted, CCC \protect\cite{bray2012}; dot-dot--dashed, relativistic 
polarized orbital \protect\cite{sullivannear}.
Experiment: squares, Stein {\it et al.}~\protect\cite{stein}; circles, 
Sinapius {\it et al.}~\protect\cite{sinapius}; diamonds, Coleman {\it et al.}~\protect\cite{coleman};
crosses, Sullivan {\it et al.}~\protect\cite{sullivannear}; triangles, Nagumo {\it et al.}~\protect\cite{nagumo_ne}.}
\label{fig:crossne}
\end{figure*}

\begin{figure*}[htp!]
\begin{center}
\includegraphics*[clip,width=0.8\textwidth]{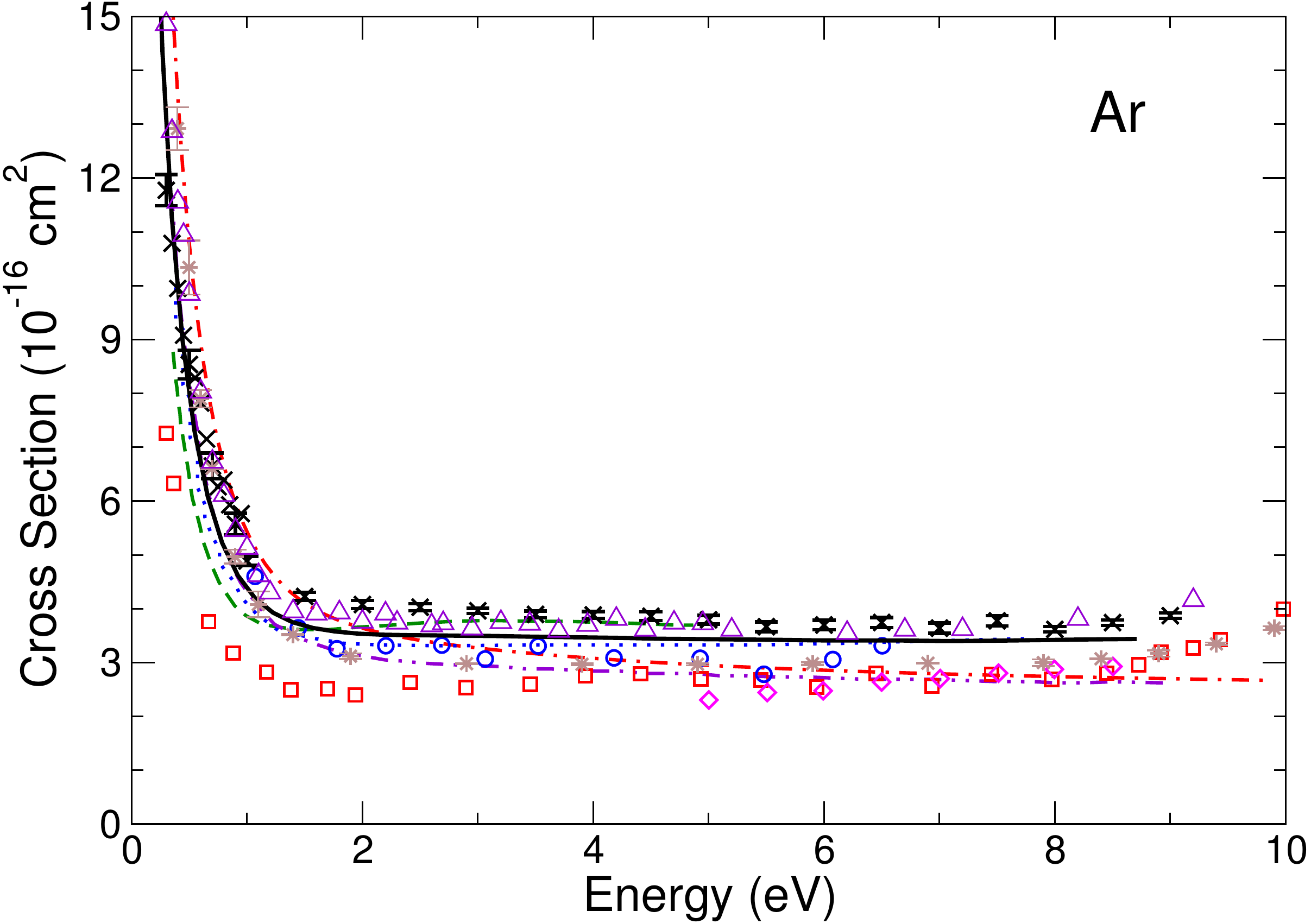}
\end{center}
\caption{Elastic scattering cross section for Ar.
Theory: solid, present many-body theory; dot-dashed, polarized orbital, \protect\cite{Mce}; 
dashed, MBPT \protect\cite{unsw2}; dotted, CCC \protect\cite{bray2012}; dot-dot-dashed, relativistic 
polarized orbital \protect\cite{sullivannear}.
Experiment: squares, Stein {\it et al.}~\protect\cite{stein1}; circles, 
Sinapius {\it et al.}~\protect\cite{sinapius}; stars, Karwasz {\it et al.}~\protect\cite{karwasz_ar}; diamonds, Coleman {\it et al.}~\protect\cite{coleman};
triangles, Zecca {\it et al.}~\protect\cite{zecca1}; crosses, Sullivan {\it et al.}~\protect\cite{sullivannear}.}
\label{fig:crossar}
%\end{figure*}
%\begin{figure*}[ht!]
\begin{center}
\includegraphics*[clip,width=0.8\textwidth]{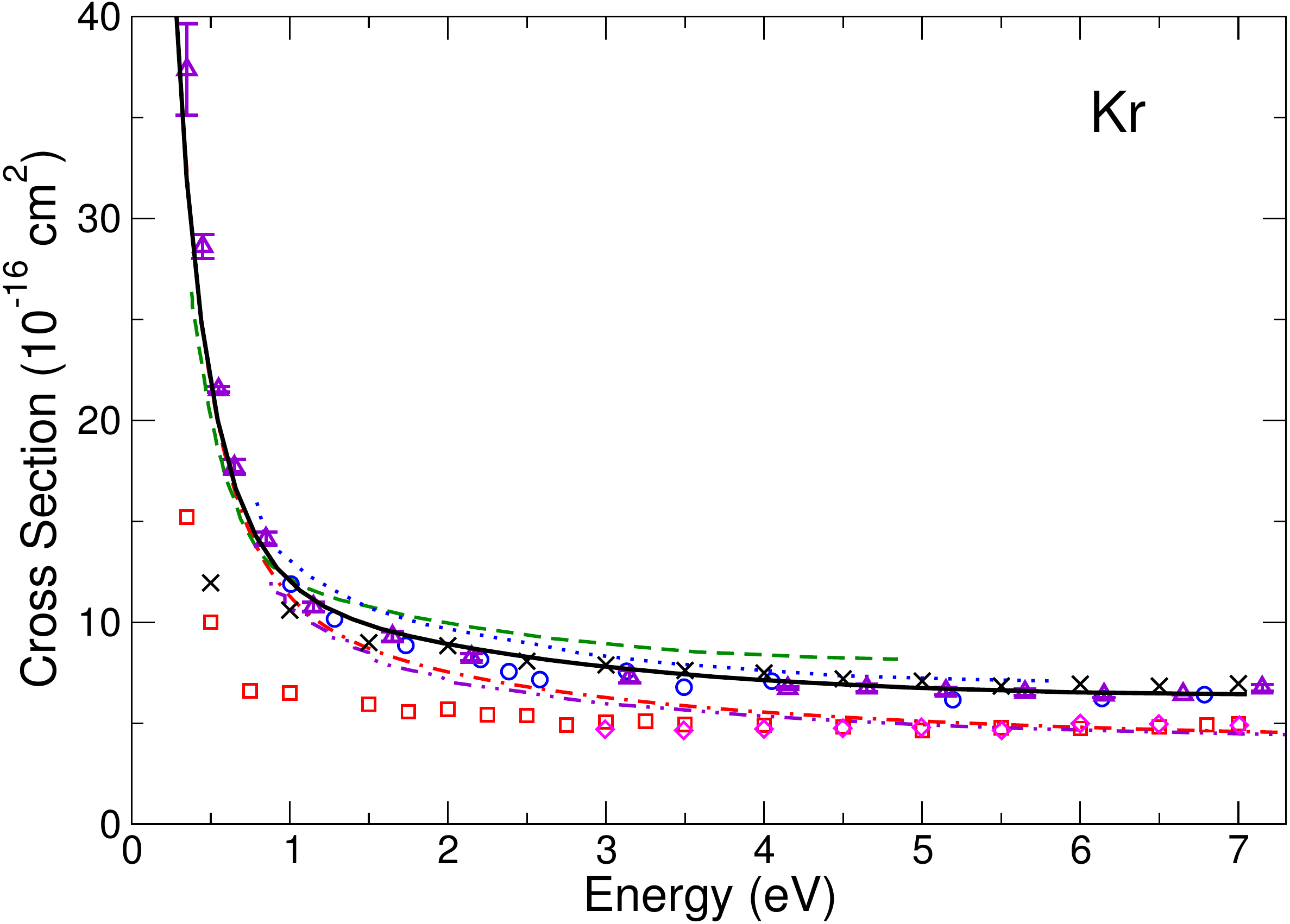}
\end{center}
\caption{Elastic scattering cross section for Kr.
Theory: solid, present many-body theory; dot-dashed, polarized orbital, \protect\cite{Mce}; 
dashed, MBPT \protect\cite{unsw2}; dotted, CCC \protect\cite{bray2012}; dot-dot-dashed, relativistic 
polarized orbital \protect\cite{sullivankr}.
Experiment: squares, Dababneh {\it et al.~}\protect\cite{dab}; circles, 
Sinapius {\it et al.}~\protect\cite{sinapius}; diamonds, Coleman {\it et al.}~\protect\cite{coleman};
triangles, Zecca {\it et al.}~\protect\cite{zecca}; crosses, Sullivan {\it et al.}~\protect\cite{sullivankr}.}
\label{fig:crosskr}
\end{figure*}

\begin{figure*}[ht!]
\begin{center}
\includegraphics*[clip,width=0.8\textwidth]{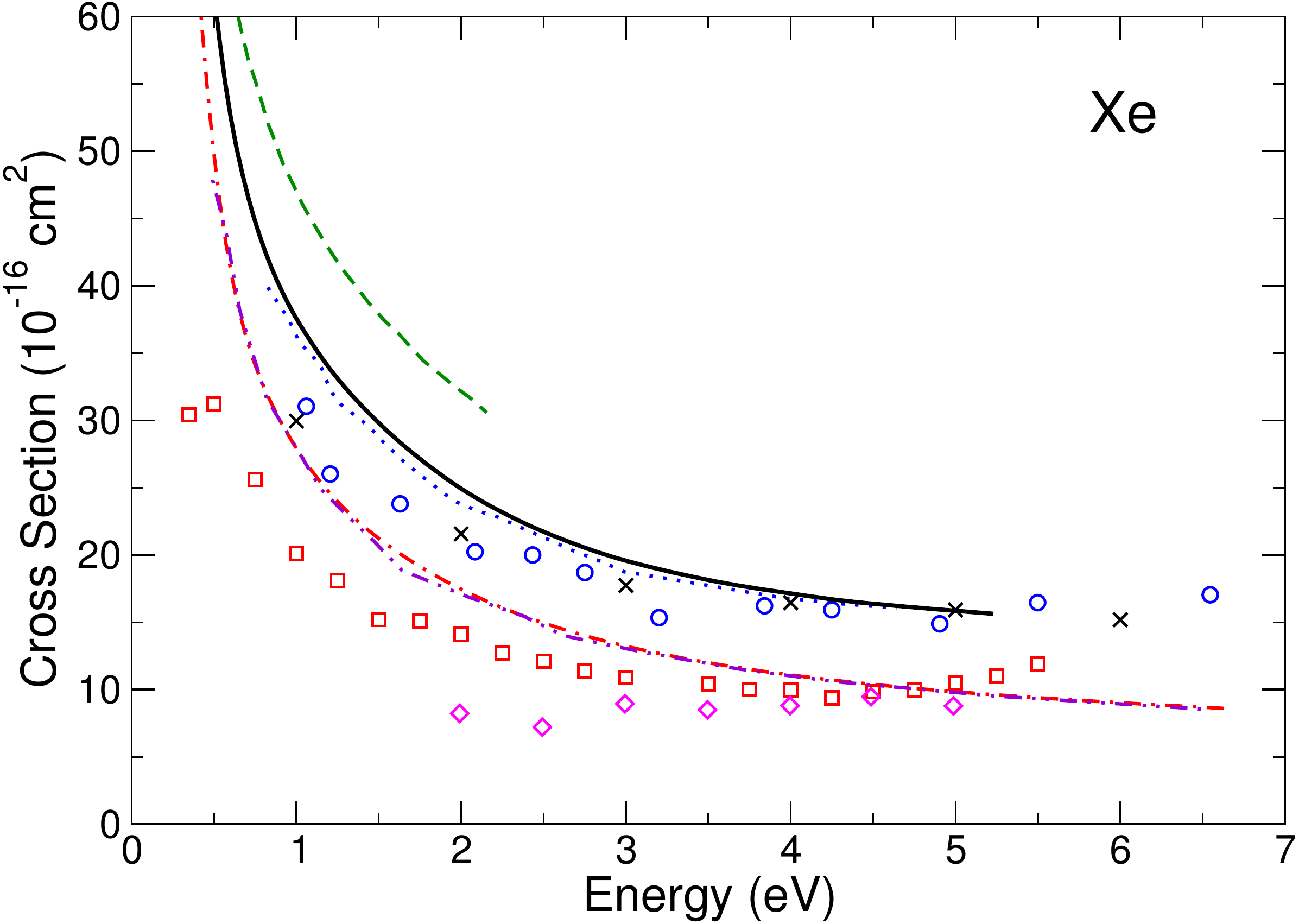}
\end{center}
\caption{Elastic scattering cross section for Xe.
Theory: solid, many-body theory; dot-dashed, polarized orbital \protect\cite{Mce}; 
dashed, Gribakin {\it et al.}~\protect\cite{unsw2}; dotted, CCC \protect\cite{bray2012}; dot-dot-dashed,
relativistic polarized orbital \protect\cite{sullivanxe}. Experiment: squares, Dababneh {\it et al.}~\protect\cite{dab}; circles, Sinapius {\it et al.}~\protect\cite{sinapius}; crosses, Sullivan {\it et al.}~\protect\cite{sullivanxe}; diamonds, Coleman {\it et al.}~\protect\cite{coleman}.\label{fig:crossxe}}
\end{figure*}

The elastic scattering cross sections along the noble gas sequence are shown in Figs.~\ref{fig:crosshe}--\ref{fig:crossxe}, where we compare them with existing experimental and theoretical data.
The numerical cross sections are tabulated in appendix \ref{app:res}.

For helium (Fig.~\ref{fig:crosshe}) our many-body theory calculations agree closely with the variational calculations \cite{humhe}, the 
convergent close-coupling results \cite{wu} and the earlier experimental measurements of Refs.~\cite{stein,mizogawa}, as well as the most recent ones \cite{sullivanhe}. A consensus on the elastic scattering cross section appears to have been reached. 
The polarized orbital results of \cite{Mce} are not in agreement with other theoretical and experimental data, and the magnetic-field-free experimental measurements of Nagumo \etal\ \cite{nagumo_he} are larger than all theory and experimental results.
%[ADD Trento and Toronto exp results. Comment on the fact that we dont see any resonant behaviour --- though look at energy resolution here --- could calculate the cross section with lots of energy points around the Trento structure to be sure].

For neon (Fig.~\ref{fig:crossne}), examining theoretical data first, the present many-body theory results agree most closely with the relativistic 
polarized orbital results \cite{sullivankr}. At energies below 2~eV, the present many-body theory results agree well
with the previous many-body theory calculations \cite{unsw2}, but trend lower
above this energy. The convergent close-coupling results of \cite{bray2012} are higher than the other calculations, while
the polarized orbital results of \cite{Mce} are considerably lower than both experiment and theory at energies above 2~eV. Comparing with experimental data, the present many-body theory results agree most closely with 
the recent measurements of Sullivan {\it et al.}~\cite{sullivannear} above 2~eV, but are lower than Sullivan's results below this energy, where they agree better with the measurements of Stein {\it et al.}~\cite{stein}.

For argon (Fig.~\ref{fig:crossar}) the many-body theory results agree well with the convergent close-coupling results
\cite{bray2012}, but are higher than the nonrelativistic and relativistic 
polarized orbital results \cite{Mce,sullivanxe} above 2~eV. Comparing to experiment, the many-body theory results are in good agreement with the measurements of Sinapius {\it et al.}~\cite{sinapius}, Zecca {\it et al.}~\cite{zecca1} and Sullivan {\it et al.}~\cite{sullivannear}, although the latter two give slightly higher values at most energies shown. Note that the Ramsauer-Townsend minimum, which is very prominent in He and Ne, is not visible in argon. This is a result of the shift of the minimum in the $s$-wave scattering cross section towards higher energies, where it gets ``filled'' with higher partial wave contributions. This interplay of the contributions of different partial waves produces a characteristic plateau in the cross section, which stretches from 2~eV to the Ps formation threshold at 8.96~eV.

For krypton, the convergent close-coupling results of \cite{bray2012} are in good agreement with the many-body theory
results, while the polarized orbital results (both relativistic \cite{Mce} and nonrelativistic \cite{sullivankr}) are lower than other theoretical predictions above 1~eV.
The experimental data of Sinapius {\it et al.}~\cite{sinapius}, and the more recent measurements by Zecca {\it et al.}~\cite{zecca} and Sullivan {\it et al.}~\cite{sullivankr} (above 2~eV) are in good agreement with each-other and with the present many-body theory results. The measurements of Dababneh {\it et al.}~\cite{dab} and Coleman {\it et al.}~\cite{coleman} lie below the many-body theory results. Compared with argon, the larger $s$-wave cross section at low energies and the increased contributions of higher partial waves produces a steadily decreasing total cross section, with no trace of the underlying Ramsauer-Townsend minimum.

For xenon, the present many-body theory results are in excellent agreement with the convergent close-coupling results
of \cite{bray2012}, lower than the previous many-body theory calculations of \cite{unsw2} (which employed an approximate treatment of virtual Ps formation), and somewhat higher than
the nonrelativistic and relativistic variants of the polarized orbital method \cite{Mce,sullivanxe}. 
Compared to experiment the present results agree most closely with the measurements of Sinapius {\it et al.}~\cite{sinapius} and Sullivan {\it et al}, especially towards higher energies. The experimental results of 
Dababneh {\it et al.~}\cite{dab} and Coleman {\it et al.~}\cite{coleman} lie below the other experimental results and the present many-body theory calculations. 
A possible reason for this may lie in the fact that the differential scattering cross section for positron scattering on heavier noble gas atoms is strongly forward peaked. Poor detection of forward-scattered positrons \cite{steindet} will cause an underestimate in the cross section.

Some general trends can be seen in the elastic scattering cross sections across the noble gas sequence. The 
many-body theory results are in good agreement with non-perturbative convergent close-coupling results \cite{bray2012}, apart from neon where there is a discrepancy that will need further investigation. 
The polarized orbital results
\cite{Mce} are seen to underestimate the cross sections at higher energies, likely due to the neglect of virtual positronium 
formation and the use of energy-independent correlation potential. Agreement with recent experimental measurements \cite{sullivanhe,zecca,zecca1,zeccaxe,sullivannear,sullivankr,sullivanxe} is generally close.

\subsection{Differential cross sections}

A quantity more sensitive to the accuracy of a scattering calculation is the differential cross section, \eqn{eq:DSC}. By making use of \eqn{phasel} for higher partial waves, the scattering amplitude (\ref{eq:f}) can be written as,
\begin{widetext}
\begin{equation}\label{scatamp1}
f(\theta)=\sum_{\ell=0}^{\ell_0}(2\ell+1)\left[\frac{e^{2i\delta_{\ell}}-1}{2ik}
-\frac{\pi\alpha _d k^2}{(2\ell-1)(2\ell+1)(2\ell+3)}\right] P_{\ell}(\cos{\theta}) - \frac{\pi\alpha _d k}{2}\sin\frac{\theta}{2} ,
\end{equation}
\end{widetext}
where $\ell_0$ is the maximum partial wave for which the phase shift has been calculated explicitly. This procedure, or some other way of effectively summing over all partial waves up to infinity, is necessary to describe the cusp of the differential cross section at $\theta =0$, which is due
to the long-range ($-\alpha _d/2r^4$) polarization potential.

\begin{figure*}[htp!]
\begin{center}
\includegraphics[clip,width=0.75\textwidth]{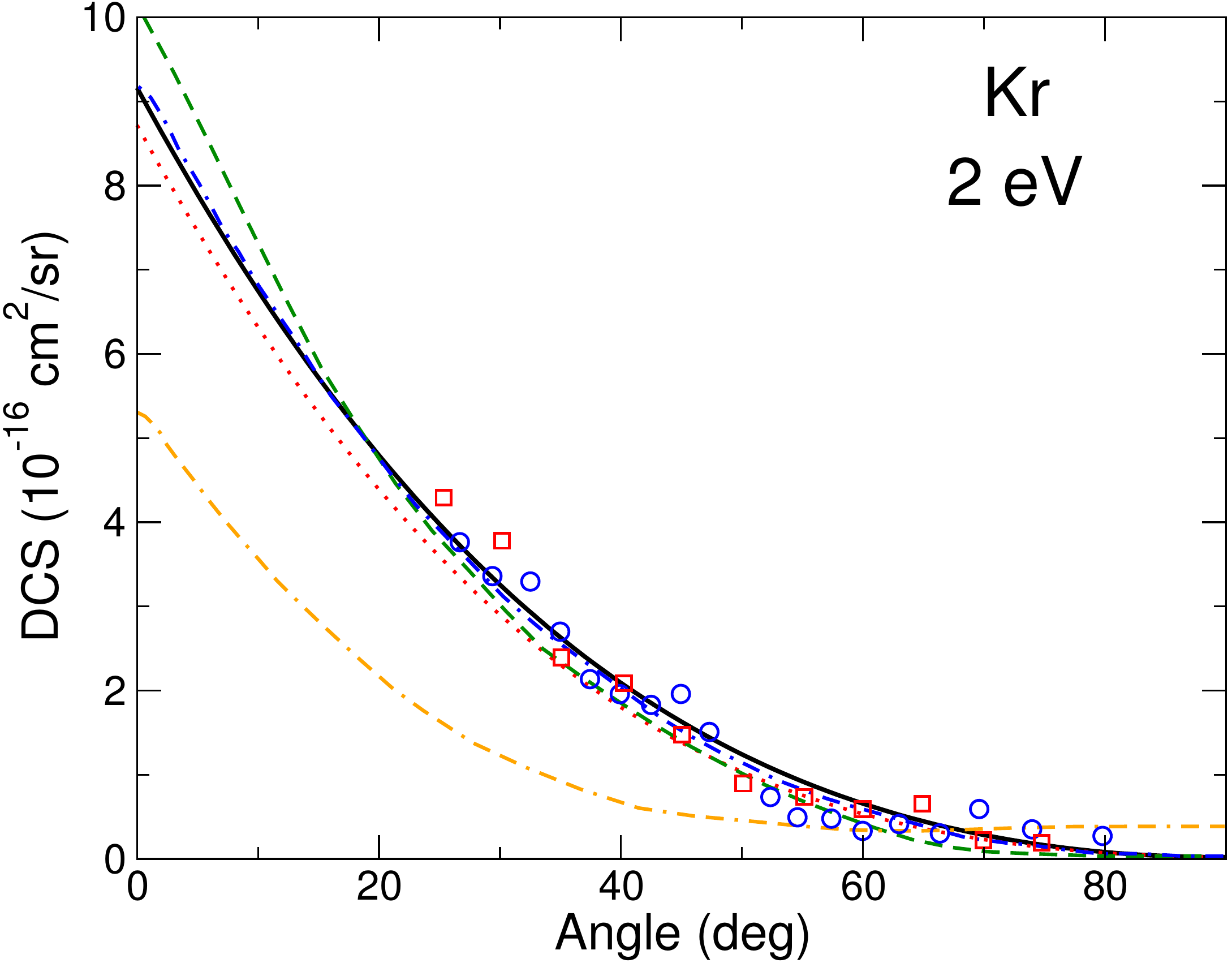}
\end{center}
\caption{Differential elastic scattering cross section for Kr.
Theory: solid, present many-body theory; dotted, polarized orbital, \protect\cite{Mce}; 
dashed, CCC \protect\cite{zecca}; dot-dashed, relativistic polarized orbital \protect\cite{sullivankr};
dashed-dashed-dotted, polarization potential \protect\cite{lam}.
Experiment: circles, Gilbert {\it et al.}~\protect\cite{gilbert}; squares, Sullivan 
{\it et al.}~\protect\cite{sullivankr}.}
\label{difcrosskr}
%\end{figure*}
%\begin{figure*}[ht!]
\begin{center}
\includegraphics[clip,width=0.75\textwidth]{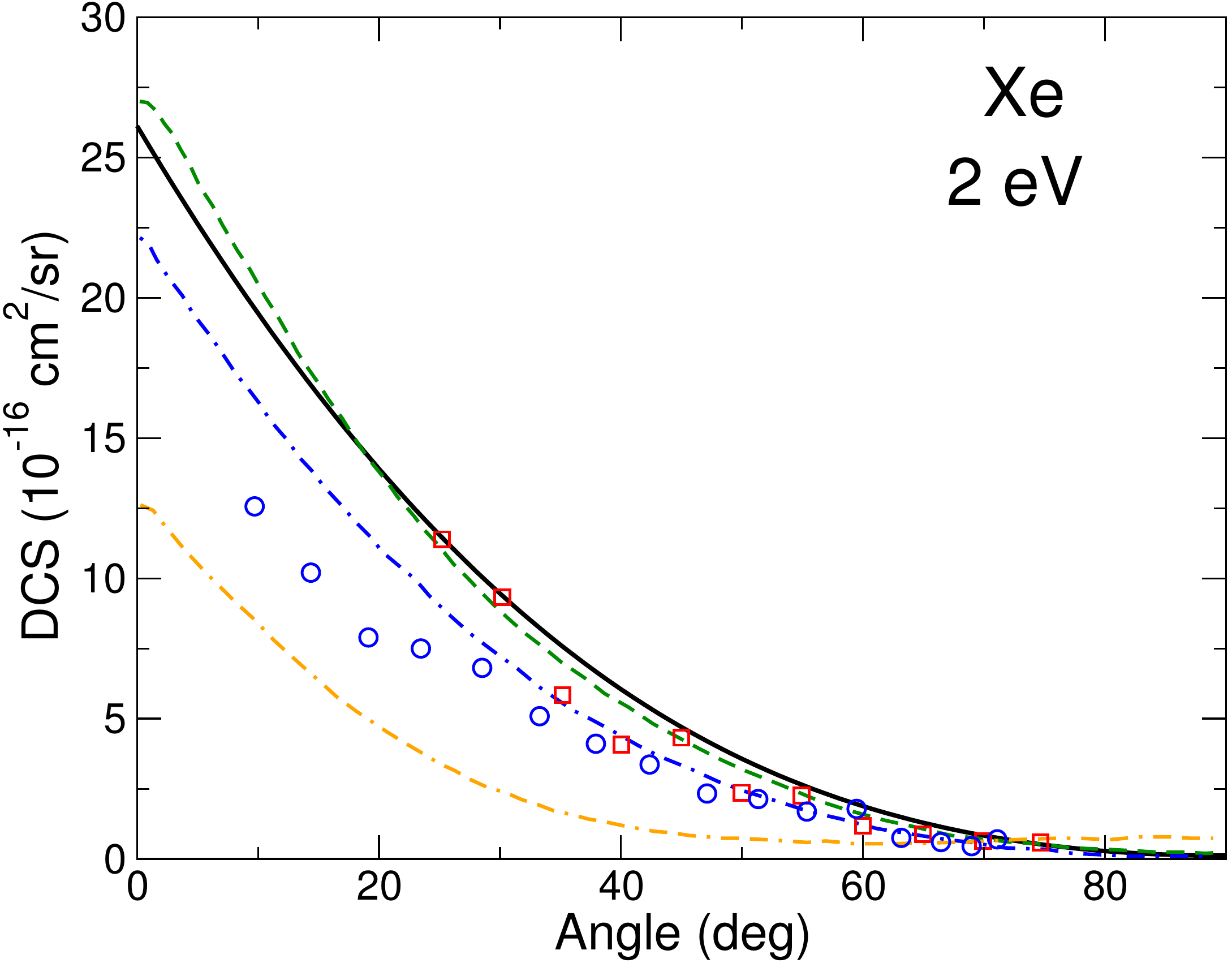}
\end{center}
\caption{Differential elastic scattering cross section for Xe.
Theory: solid, present many-body theory; 
dashed, CCC \protect\cite{sullivanxe}; dot-dashed, relativistic polarized orbital
\protect\cite{sullivanxe};
dashed-dashed-dotted, polarization potential \protect\cite{lam}.
Experiment: circles, Marler {\it et al.}~\protect\cite{marler}; squares, Sullivan 
{\it et al.}~\protect\cite{sullivanxe}.}
\label{difcrossxe}
\end{figure*}

Figures \ref{difcrosskr} and \ref{difcrossxe} compare differential cross sections calculated using our many-body 
phase shifts for krypton and xenon at an incident positron energy of 2~eV with other theoretical calculations and 
experimental data from the San Diego \cite{gilbert, marler} and ANU groups \cite{sullivankr,sullivanxe}. 
It is important to note that these groups use positron traps to accumulate positrons, which are then extracted to form a pulsed, energy tunable positron beam. This beam is magnetically guided to the target gas cell, and the differential cross section is measured by observing the change in the longitudinal positron energy. However, the apparatus is not able to distinguish between forward-scattered and back-scattered particles, 
as any back-scattered particles are reflected and passed back through the gas cell. The measured differential cross section for a scattering angle $\theta$ is therefore the sum of the cross sections at the angle $\theta$ and $180-\theta$. Theoretical results have therefore been folded about 90$^\circ$ where necessary.

For krypton (Fig.~\ref{difcrosskr}) the present many-body theory calculations are in good agreement with the CCC results \cite{zecca}, nonrelativistic and relativistic variants of the polarized orbital method \cite{Mce,sullivankr}, and the experimental data of \cite{gilbert} and \cite{sullivankr} across the angular range, with the CCC values trending
slightly higher at small angles. (The latter is compatible with the larger absolute value of the scattering length in the CCC calculation, Table~\ref{stpol}.) However, a large discrepancy is observed with the polarization calculations of \cite{lam}. In the case of xenon (Fig.~\ref{difcrossxe}), there is excellent agreement between the present many-body calculations and the CCC calculations \cite{sullivanxe}, with both calculations being somewhat higher than the relativistic polarized orbital calculations
\cite{sullivanxe}. As in krypton, the present results show a large difference with the polarization potential calculations of \cite{lam}. The many-body theory and CCC calculations are in better agreement with the ANU experiment
\cite{sullivanxe} than the data of Ref.~\cite{marler}, with the two experiments in better accord above 40$^{\circ}$. The stronger peaking of both many-body theory and CCC results at $\theta =0$ can be related to the larger values of the scattering lengths in these calculations compared to the polarized-orbital result (Table~\ref{stpol}). The large scattering length in Xe is behind the strongly enhanced positron annihilation rates at low (e.g., room-temperature, thermal) energies in xenon (see Sec.~\ref{sec:ann}). We, therefore, believe that the many-body theory, CCC and ANU data 
are more accurate for $\theta <30^\circ $.

%\newpage
\section{Positron annihilation}\label{sec:ann}
 
\subsection{Energy resolved $\Zeff$}

\begin{figure*}[ht!]
\begin{center}
\includegraphics*[clip,width=0.452\textwidth]{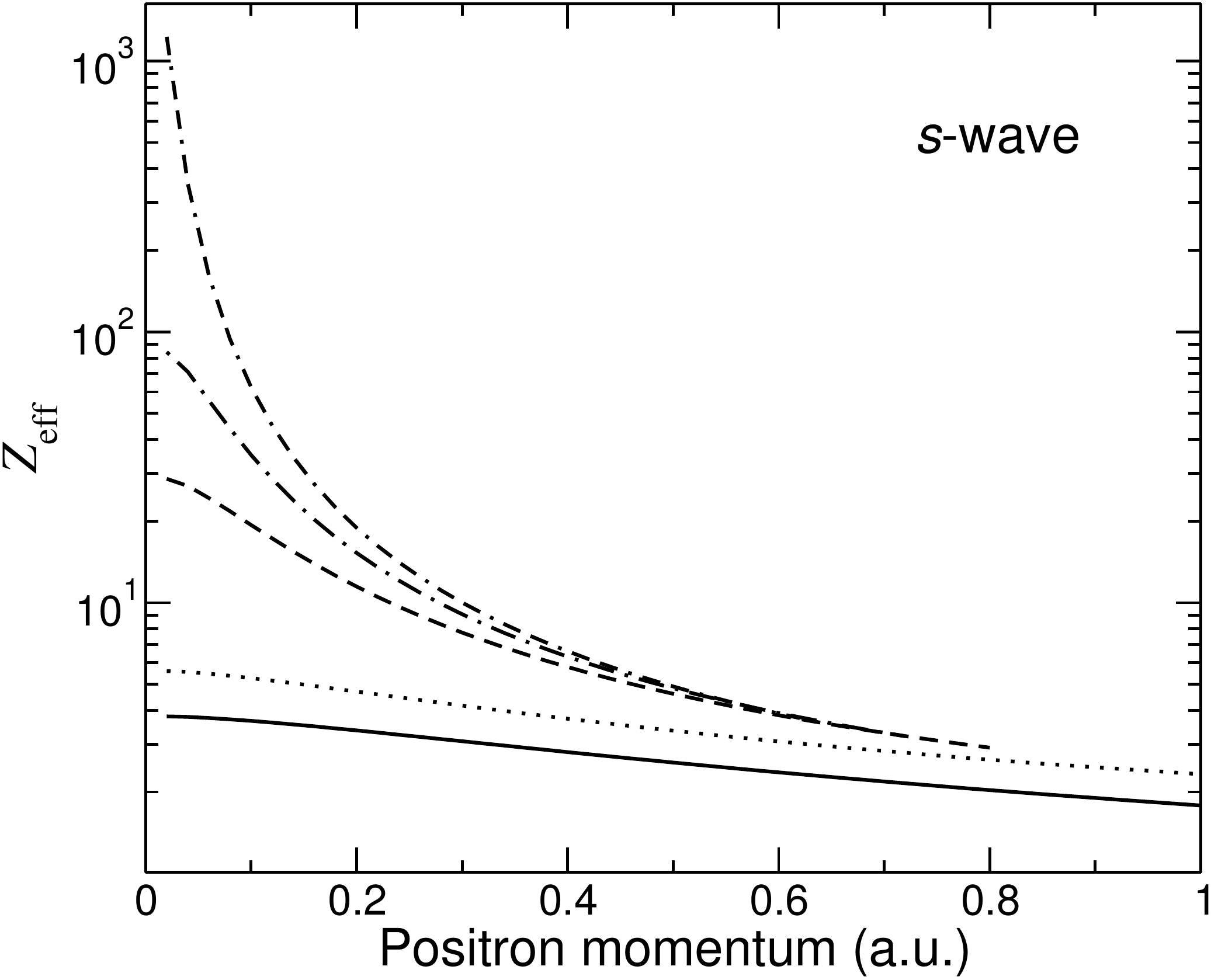}

\includegraphics*[clip,width=0.452\textwidth]{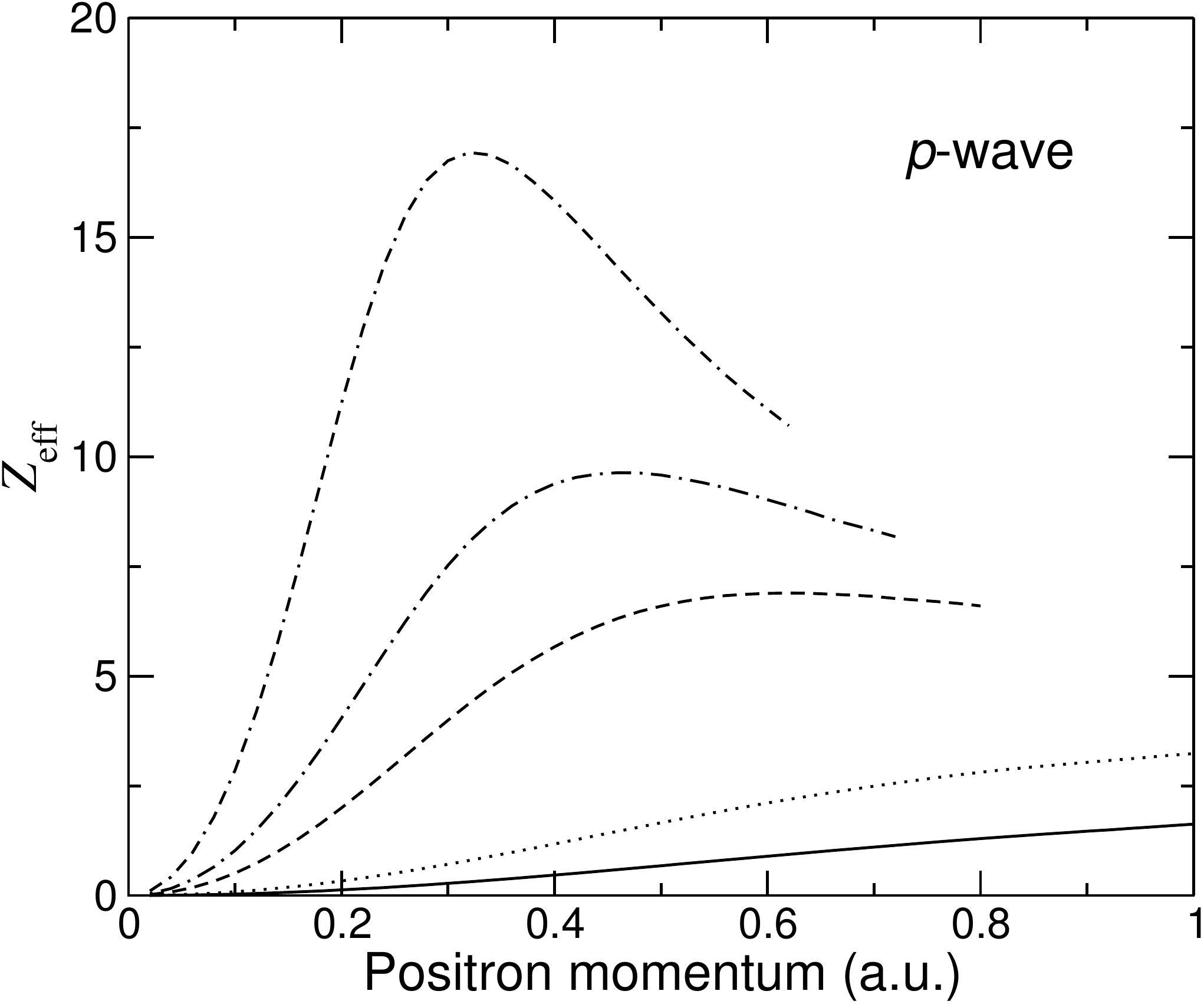}

\includegraphics*[clip,width=0.452\textwidth]{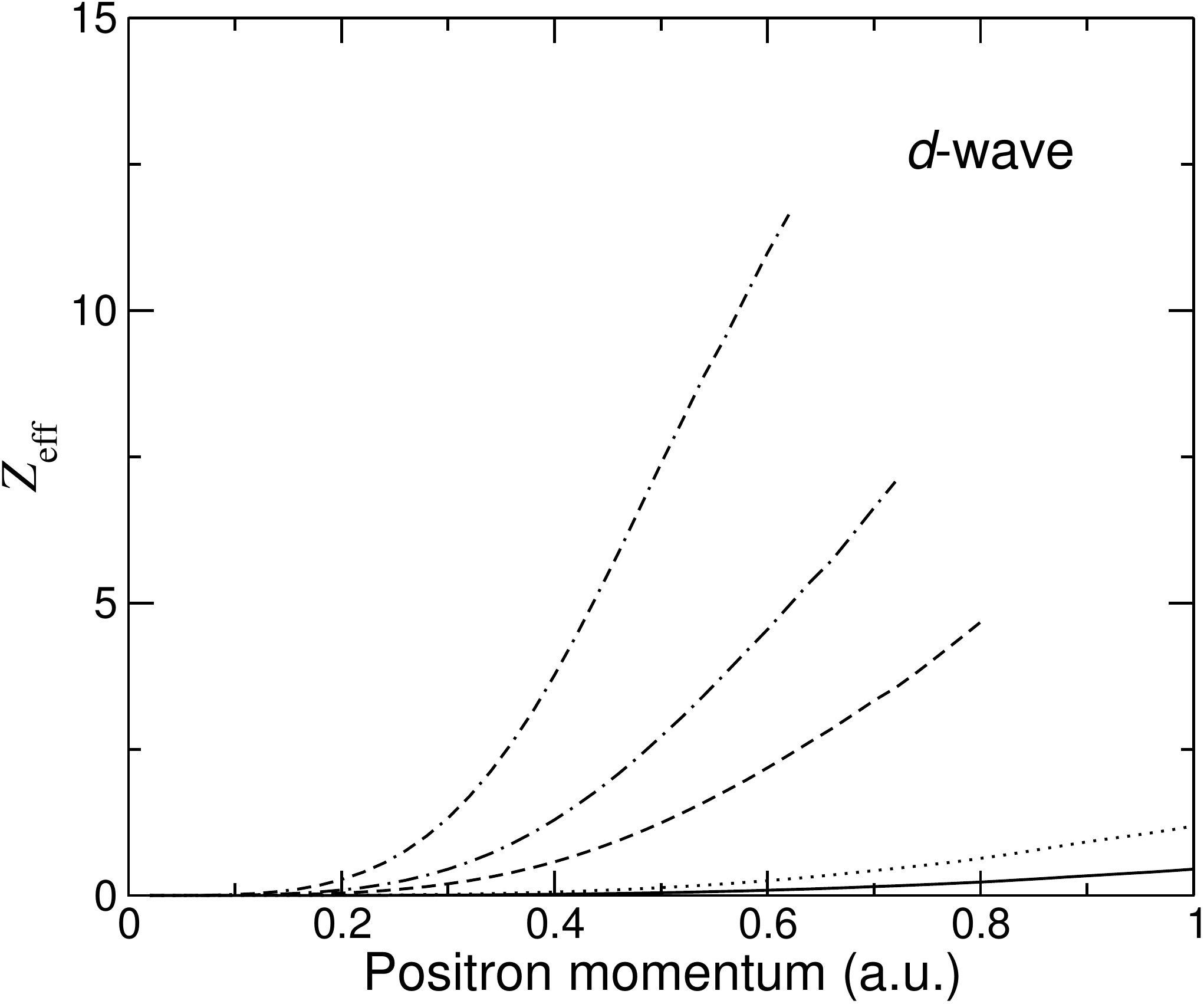}
\end{center}
\caption{Contributions of the $s$-, $p$- and $d$-waves to the annihilation rate $\Zeff$ on the valence subshells of the noble gases: helium ($1s$, solid line); neon ($2s+2p$, dotted line); argon ($3s+3p$, dashed line); krypton ($4s+4p$, dash-dotted line) and 
xenon ($5s+5p$, dash-dash-dotted line).}
\label{zeffpartial}
\end{figure*}

Figure \ref{zeffpartial} shows partial-wave contributions to $\Zeff$ for positron annihilation on the valence shell electrons of the noble gas atoms as functions of the positron momentum. Note that at low positron momenta $k$
the Wigner threshold law predicts $\Zeff \propto k^{2\ell}$ for the positron with the orbital angular momentum $\ell $ \cite{lan}. As a result, the $s$-wave contribution dominates at low energies.

As one moves along the noble gas sequence, the $s$-wave $\Zeff$ becomes increasingly large and strongly peaked at low energies. This strong energy dependence is due to the existence of positron-atom virtual levels \cite{GS64,Dzuba,unsw2}, which is signified by large scattering lengths $a$
for Ar, Kr and Xe. In this case the momentum dependence of $\Zeff$ at low energies can be described analytically \cite{Dzuba,unsw2,GG00,NewDir,MI02}, as
\begin{equation}\label{fit}
\Zeff(k)=\frac{K}{\kappa^2+k^2} + A,
\end{equation}
where $K$ and $A$ are constants. The first term in \eqn{fit} is due to
the $s$-wave contribution enhanced by the virtual label (i.e., small
$\kappa =1/a$, $|\kappa |\ll 1/R_a\sim 1$~a.u., where $R_a$ is the atomic radius). The constant term $A$ accounts for the nonresonant background and contributions of higher partial waves to $\Zeff$. Equation (\ref{fit})
makes it clear that if the thermal positron momentum $k$ is smaller than $\kappa$ then the annihilation rate will be proportional to the scattering length squared.
  
The $p$-wave $\Zeff$ also appears to show the formation of a broad shape resonance especially prominent for Xe. This is supported by the behaviour of the $p$-wave phase shift, e.g., that for Kr shown in \fig{phasekr}. For more polarizable targets which generate stronger positron attraction, such as Mg and Zn, $p$-wave resonances in $\Zeff$ become a prominent feature of the energy dependence of the annihilation  parameter \cite{MZ08}.

\subsection{Thermally averaged $\Zeff$}

Most of the available experimental data for $\Zeff$ in noble gases have been obtained for thermalized positrons at room temperature \cite{MB04}. As can be seen from \fig{zeffpartial}, the annihilation rates in helium and neon have a weak energy dependence in the range of thermal positron momenta ($k\sim 0.045$~a.u.~at room temperature). Hence, for these atoms we take the $\Zeff$ values at $k=0.04$~a.u.~to compare with experiment (see Table~\ref{stpol1}). The
$\Zeff$ values at such low momenta are primarily due to the $s$-wave positron annihilation, with the $p$-wave contributing only a fraction of one per cent. In addition, for neon about 0.3\% is due to positron annihilation with the core ($1s$) electrons (Table \ref{app1b}).

\begin{table}[ht!]
\begin{ruledtabular}
\caption{Thermally averaged $\Zeff$ for the noble gases.}
\label{stpol1}
\begin{tabular}{cccccc}
 & He & Ne & Ar & Kr & Xe\\
\hline
$\bar{Z}_{\rm eff}$\footnote{Present many-body theory calculations.} & 3.79 &5.58 &26.0 &66.1 &450\\
$\bar{Z}_{\rm eff}$\footnote{He: Kohn variational calculations \cite{humheannih}; Ne, Ar, Kr, Xe: polarized orbital calculations \cite{Mce}.} & 3.88 &6.98 &30.5 &56.3 &202\\
$\bar{Z}_{\rm eff}$\footnote{Experiment: He, Ne, Ar \cite{coleman75}; Kr \cite{charlton}, Xe \cite{griffith,charlton}.} & 3.94 & 5.99 & 26.77 & 65.7 & 320, 400--450\\ 
$\bar{Z}_{\rm eff}$\footnote{Experiment: Ar and Kr \cite{iwata1}; Xe \cite{surko}.} & -- & -- &33.8 &90.1 & 401
\end{tabular}
\end{ruledtabular}
\end{table}

\begin{figure*}[ht!]
\begin{center}
\includegraphics[clip,width=0.8\textwidth]{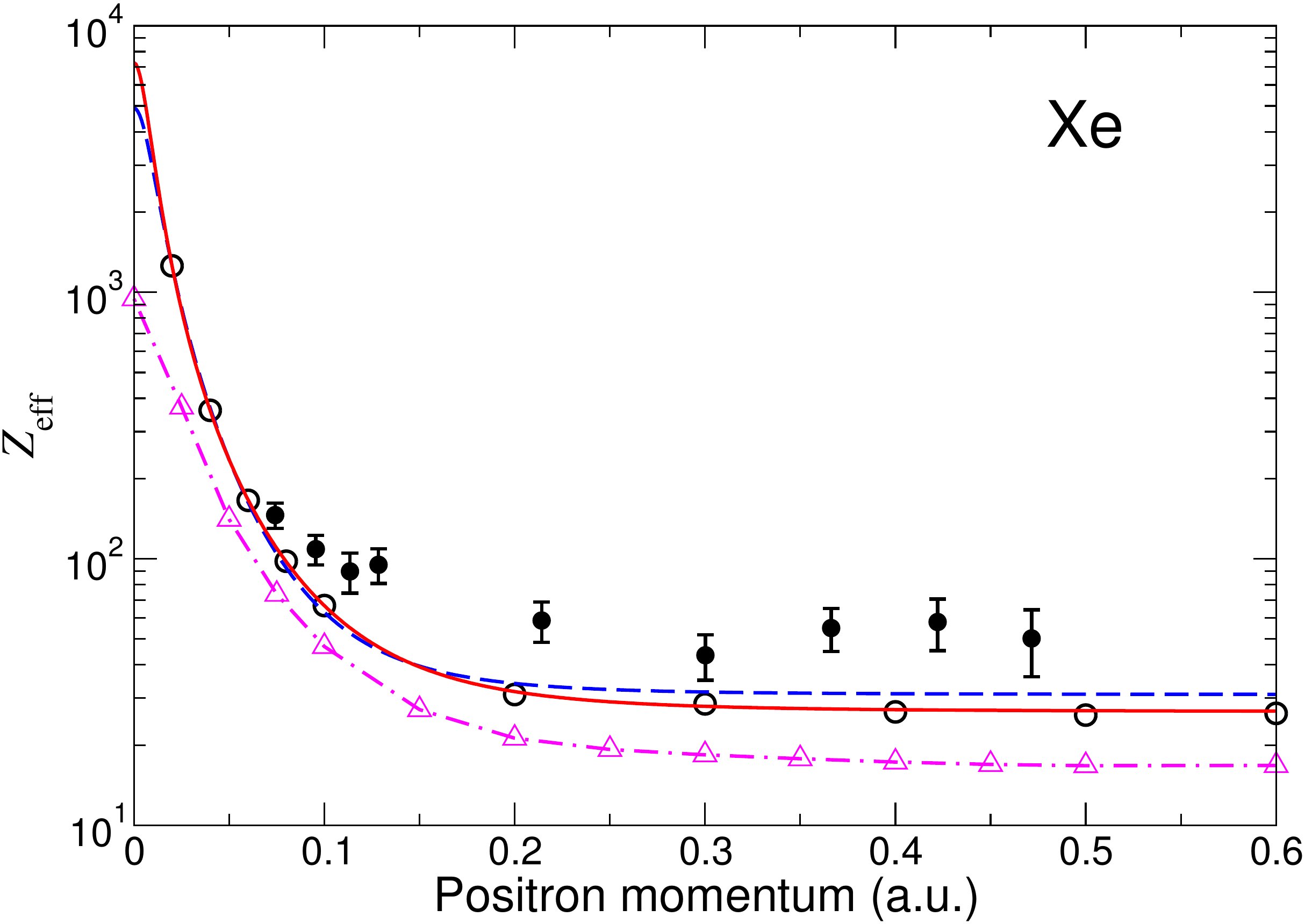}
\end{center}
\caption{Calculated $\Zeff $ for Xe (sum of $s$-, $p$- and $d$-wave values including both valence and core contributions, open circles) and their fits (\ref{eq:fit}), in which $\kappa =1/a =-0.0118$~a.u.~is fixed (dashed line), or used as a fitting parameter (solid line). Solid circles are experimental values from Ref.~\cite{MB04}. Triangles with dot-dashed line are the polarized orbital results \cite{Mce}.}
\label{fig:xezeff_fit}
\end{figure*}

For argon, krypton and xenon the energy dependence of $\Zeff$ becomes progressively stronger and we compute thermal $\Zeff$ values from the Maxwellian average
\begin{equation}\label{eq:ZeffT}
\bar{Z}_{\rm eff}=\int_0^{\infty} \Zeff(k)
\frac{\exp(-k^2/2k_BT)}{(2\pi k_BT)^{3/2}}4\pi k^2 dk ,
\end{equation}
where $k_B$ is the Boltzmann constant and $k_BT=9.29\times 10^{-4}$~a.u.
at room temperature $T=293$~K. The integration is performed using a fit of the calculated total $\Zeff$ values ($s$-, $p$-, and $d$-wave, valence and core), of the form
\begin{equation}\label{eq:fit}
\Zeff(k)=\frac{K}{\kappa^2+k^2+\beta k^4} + A,
\end{equation}
over the momentum range $k=0.02$--0.3~a.u. This fit is based on \eqn{fit}, with an extra $k^4$ term included to improve the accuracy. This is especially important for xenon where $\Zeff$ has the most vigorous momentum dependence. Figure \ref{fig:xezeff_fit} shows the
calculated $\Zeff$ for Xe (open circles) \cite{comm:Zeff}, together with two fits of the form (\ref{eq:fit}), experimental data from Ref.~\cite{MB04}, and $\Zeff$ from the polarized orbital calculations \cite{Mce}.

In the first fit shown in \fig{fig:xezeff_fit} by the dashed line, the parameter $\kappa $ is fixed by the value of the scattering length, $\kappa =1/a =-0.0118$~a.u., and the values of the other parameters are $K=0.683$, $A=30.95$ and $\beta =111.8$~a.u. This fit gives $\bar Z_{\rm eff} =448$ at $T=293$~K. In the second fit shown by the solid line, $\kappa $ is regarded as a free parameter. The corresponding set of values $K=0.6047$, $\kappa =-0.00914$, $A=26.74$, and $\beta =50.49$~a.u.~produces an excellent fit of the numerical $\Zeff$ over the whole momentum range. It gives $\bar Z_{\rm eff} =458$. Given the 2\% difference between the two values, we report $\bar Z_{\rm eff} =450$ as our best prediction of the room-temperature value for Xe.

Using \eqn{eq:fit} as a four-parameter fit for Ar ($K=0.2002$, $\kappa =-0.1048$, $A=11.26$, and $\beta =21.68$~a.u.) yields $\bar Z_{\rm eff} =26.0$ at $T=293$~K. The fit for Kr ($K=0.3659$, $\kappa =-0.0696$, $A=16.04$, and $\beta =30.74$~a.u.) produces $\bar Z_{\rm eff} =66.1$. Although the momentum dependence of $\Zeff$ in Ar and Kr is not nearly as steep as in Xe, thermal averaging is important for them. For example, the calculated values at $k=0.04$~a.u.~are $\Zeff =27.1$ for Ar and 72.1 for Kr, while the thermally averaged values obtained above are close to $\Zeff (k)$ at $k\approx 0.048$~a.u.

The present many-body theory and the $\bar{Z}_{\rm eff}$ values obtained above are a significant improvement on the previous many-body theory study \cite{unsw2}. In Table~\ref{stpol1} the calculated $\bar{Z}_{\rm eff}$ values for the noble gas atoms are compared with available experimental data and some theory values. For helium the many-body theory result is in good agreement with precise variational calculations \cite{humheannih} and with the measurements \cite{coleman75}, the discrepancy being less than 5\%. For neon, argon and krypton the many-body results are in close agreement with the earlier measurements \cite{coleman75}, but differ significantly
from the positron-trap results for Ar and Kr \cite{iwata1}, and are lower than the polarized orbital calculations \cite{Mce}. Even assuming an error of 5--10\% in the many-body calculations, the results of Iwata \etal \cite{iwata1} for Ar and Kr appear to be anomalously high. For xenon, the many-body result is much higher than that from the polarized orbital calculations \cite{Mce} and is in reasonable accord with the experiment of \cite{surko} and one set of earlier measurements \cite{charlton}. Given the approximations made in the polarized orbital method, the discrepancy between its results and our theory are not unexpected. (In fact it is remarkable that this relatively crude method produces such reasonable results.) To gain a better understanding of the discrepancies between theory and experiment, it is necessary to examine the experimental techniques used to measure $\bar{Z}_{\rm eff}$.

The results of Refs. \cite{coleman75,charlton,griffith} were obtained using positron lifetime spectroscopy. In this technique, positrons from a radioactive source (e.g., $^{22}$Na) are injected into a gas cell, where they thermalize and annihilate. The lifetime is measured as the delay between  
the nuclear gamma ray emitted in the $\beta^+$ decay, and the annihilation
gamma ray. By measuring the lifetime of $10^6-10^7$ positrons it is possible to obtain a lifetime spectrum. The annihilation rate can then be found from the exponential fit of this spectrum. For lighter nobles gases (He to Kr) thermalization occurs much faster than annihilation, which provides for a reliable measurement of $\bar{Z}_{\rm eff}$ with fully thermalized positrons. In xenon the positron energy loss in momentum-transfer collisions is smaller due to the large mass of the atom, while the $\Zeff$ values in Xe are higher. As a result, the annihilation rate is measured for epithermal positrons, resulting in $\Zeff$ values lower than expected (e.g., $\bar{Z}_{\rm eff}=320$ \cite{griffith}). Adding small amounts of a lighter, low-$\Zeff$ gas, e.g., He or H$_2$, to Xe allows for fast thermalization and produces truly thermalized annihilation rates with $\bar{Z}_{\rm eff}=400$--450 \cite{charlton}. As seen in Table~\ref{stpol1}, these values are in good agreement with the present calculation.

The results of Refs. \cite{surko,iwata1} were obtained using a Penning-Malmberg positron trap. In this type of experiment positrons from a radioactive source are slowed down to electron-volt energies using a solid neon moderator. They are then accelerated towards the trap region,
and become trapped by loosing energy through inelastic collisions with a buffer gas, such as N$_2$. Differential pumping of the buffer gas allows the 
thermal positrons to be stored for a long time in a high-vacuum region of the trap. A test gas is then injected into the trap at a known low pressure, and the positron annihilation rate is found by determining the number of remaining positrons as a function of time. In this set-up positrons are well thermalized, and the result for Xe, $\bar{Z}_{\rm eff}=401$ \cite{surko}
is broadly consistent with that of Ref.~\cite{charlton} and our calculation.
The current calculations and available experimental data thus indicate that the value of $\bar{Z}_{\rm eff}$ for xenon lies in the range 400--450.
However, the discrepancy between the present results and those of Ref.~\cite{iwata1} for Ar and Kr is of concern, especially given the agreement we between our $\bar{Z}_{\rm eff}$ and the earlier gas-cell measurements \cite{coleman,charlton}. It would be worthwhile for new positron-trap measurements of $\bar{Z}_{\rm eff}$ for argon and krypton to be performed to help resolve this discrepancy.

%\newpage
\section{Conclusions}

The many-body formalism that has been developed to study positron scattering and annihilation from atoms and ions \cite{hydrogen,spectra,halogen} has been applied to the noble gas sequence. 
Good agreement with experimental data and recent non-perturbative CCC calculations has been obtained for the elastic scattering 
cross sections. Calculated thermally averaged $\Zeff$ for He, Ne, Ar, and Kr are in good agreement with experimental values obtained using positron lifetime spectroscopy in gas cells. For xenon theoretical and experimental data point to a thermal room-temperature $\bar{Z}_{\rm eff}=400$--450. Theoretically, the uncertainty is related to the difficulty in predicting the exact value of the large positron-xenon scattering length. Experimentally, the gas-cell measurements are affected by slow positron thermalization in xenon and the need to use gas mixtures to achieve thermalization. The more sophisticated positron-trap set up allows measurements of energy-resolved $\Zeff$. The corresponding results for Xe (and Ar \cite{MB04}) are generally in accord with the calculation, although slightly higher. Being more difficult, trap-based measurements of absolute $\Zeff$ may suffer from larger systematic errors. In addition, in systems with rapidly varying $\Zeff $, such as Xe, the energy-resolved low-energy ($\lesssim 0.1$~eV) data can be affected by the positron-beam energy distribution, so a more detailed comparison is required.

Through this work, together with Ref.~\cite{DGGinnershells}, the many-body theory method has provided a near complete understanding of the positron-noble gas atom system at positron energies below the Ps-formation threshold.
Positron-scattering phase shifts and cross sections, and rates and $\gamma$-spectra for positron annihilation on core and valence electrons, have been calculated in a consistent framework that takes proper account of positron-atom and positron-electron correlations.
There are, however, a number of ways in which the many-body theory can be developed and extended. Thus, it should be straightforward to generalize our many-body theory to a fully relativistic formalism. This will be important in exploring the influence that relativistic effects have on positron scattering from high-$Z$ atoms and ions, particularly for positron annihilation on inner shells. It will also be important to account for higher-order polarization effects \cite{relat1, relat2} beyond the third order many-body diagrams included in the present work, particularly for xenon where the low-energy cross section is highly sensitive to correlation effects. Another area of interest is the application of the many-body theory to open-shell systems.  
Although only truly rigorous for closed-shell systems, approximate methods can be introduced that should allow a reasonably accurate application of the theory to such systems. 
This would be particularly useful for studies of positron annihilation on core electrons in condensed matter systems~\cite{condense1,DGGinnershells}. Finally, the understanding gained by studying positron-atom interactions is very valuable for gaining insights into positron-molecule interactions. The latter is a much more complex system, in which positron binding and resonances, as well as the vibrational dynamics of the positron-molecule complex, play a crucial role in providing strongly enhanced annihilation rates \cite{RMP}.

\acknowledgments
DGG is grateful to the Institute for Theoretical Atomic, Molecular and Optical Physics, at the Harvard-Smithsonian Centre for Astrophysics (Cambridge, MA, USA),
where he carried out part of this work as a visitor, and is indebted to H.~R.~Sadeghpour and colleagues for their generous hospitality. 
DGG was supported by DEL Northern Ireland.

\appendix

%\newpage
%\begin{widetext}
\onecolumngrid

\section{Algebraic form of the many-body diagrams}\label{app:diag}

The algebraic expressions for the various many-body diagrams are tabulated
below and complement those provided in Appendix A of Ref.~\cite{hydrogen}. The derivation of these expressions makes extensive use of graphical
techniques for performing angular momentum algebra \cite{vars}.
 
Let us denote the direct reduced Coulomb matrix element by
\begin{align}
\langle 3,4\|V_l\|2,1\rangle &=\sqrt{[l_3][l_4][l_2][l_1]}
\tj{l_1}{l}{l_3}{0}{0}{0}
\tj{l_2}{l}{l_4}{0}{0}{0}\nonumber \\ \label{dir}
&\times \int\!\!\int\!\!P_{\eps_3l_3}(r_1)P_{\eps_4l_4}(r_2)
\frac{r_{<}^{l}}{r_{>}^{l+1}}P_{\eps_2l_2}(r_2)P_{\eps_1l_1}(r_1)
dr_1dr_2,
\end{align}
where $[l]\equiv 2l+1$.
Secondly, we denote the reduced Coulomb matrix element for an electron-positron pair with the total angular momentum $J$ by
\begin{eqnarray}\label{exch}
\langle 3,4\|V^{(J)}\|1,2\rangle
&=&\sum_l (-1)^{J+l}\sj{J}{l_3}{l_4}{l}{l_2}{l_1}
\langle 3,4\|V_l\|2,1\rangle.
\end{eqnarray}
Note that this is similar to the `exchange' matrix elements that one meets 
in electron scattering \cite{amusiaelect,Amusia2,amusia75}. 
These matrix elements form the main components of the analytic expressions represented by the diagrams. The additional rules are as follows.
For closed-shell atoms, each electron-hole `loop' gives a spin factor of 2.
In addition, asymmetric $\Zeff$ diagrams should contain an extra factor of
2 arising from their mirror images. The sign of each diagram is given by
$(-1)^{a+b+c}$, where $a$ is the number of hole lines, $b$ is the number of `loops' and $c$ is the number of positron-electron interactions.

Matrix elements of the $\delta$-function operator are defined similarly to those of the Coulomb interaction. We denote the direct matrix element by
\begin{align}
%\begin{split} %places Eqn number between the two lines
\langle 3,4\|\delta_l \|2,1\rangle &=\frac{[l]}{4\pi }\sqrt{[l_3][l_4][l_2][l_1]}
\tj{l_1}{l}{l_3}{0}{0}{0}
\tj{l_2}{l}{l_4}{0}{0}{0}\nonumber \\ \label{dirz}
&\times \int P_{\eps_3l_3}(r)P_{\eps_4l_4}(r)
\frac{1}{r^2}P_{\eps_2l_2}(r)P_{\eps_1l_1}(r)dr,
%\end{split}
\end{align}
and the matrix element for a positron-electron pair coupled to an angular momentum $J$ by
%will be denoted by,
\begin{eqnarray}\label{exchz}
\langle 3,4\|\delta^{(J)}|
|1,2\rangle =\sum_l (-1)^{J+l}\sj{J}{l_3}{l_4}{l}{l_2}{l_1}
\langle 3,4\|\delta^{(J)}\|2,1\rangle.
\end{eqnarray}

The algebraic expressions for the third-order contributions to $\langle \eps '|\Sigma _E|\eps \rangle $ shown by the diagrams in \fig{fig:Sig_3}\,(a)--(e) 
are, respectively:
%\begin{widetext}
% this is the 3rd order self-energy diagram with electron-hole attraction.
\begin{equation}\label{firss}
%{\rm Figure~\ref{fig:Sig_3}(a)}:~~~~~~-
-\sum_{\stackrel{\nu,\mu_1,\mu_2>F}{n_1,n_2\leq F}}\sum_l
\frac{2\langle \eps',n_2\|V_l\|\mu_2,\nu\rangle
\langle \mu_2,n_1\|V^{(l)}\|\mu_1,n_2\rangle
\langle \nu,\mu_1\|V_l\|n_1,\eps\rangle}
{(2l+1)(2\ell +1)(E+\eps_{n_2}-\eps_{\nu}-\eps_{\mu_2})
(E+\eps_{n_1}-\eps_{\nu}-\eps_{\mu_1})},
\end{equation}
%Figure \ref{fig:Sig_3}(b):
% this is a 3rd order self-energy diagram.
\begin{equation}\label{firs3}
%{\rm Figure~\ref{fig:Sig_3}(b)}:~~~~~~-
\sum_{\stackrel{\nu,\mu_1,\mu_2>F}{n_1,n_2\leq F}}\sum_l
\frac{4\langle \eps',n_2\|V_l\|\mu_2,\nu\rangle
\langle n_1,\mu_2\|V_l\|n_2,\mu_1\rangle
\langle \nu,\mu_1\|V_l\|n_1,\eps\rangle}
{(2l+1)^2(2\ell +1)(E+\eps_{n_2}-\eps_{\nu}-\eps_{\mu_2})
(E+\eps_{n_1}-\eps_{\nu}-\eps_{\mu_1})},
\end{equation}
%Figure \ref{fig:Sig_3}(c):
% this is the 3rd order self-energy diagram with simultaneous double excitation.
\begin{equation}\label{firs2}
%{\rm Figure~\ref{fig:Sig_3}(c)}:~~~~~~-2
\sum_{\stackrel{\nu,\mu_1,\mu_2>F}{n_1,n_2\leq F}}\sum_l
\frac{4\langle \eps',n_2\|V_l\|\mu_2,\nu\rangle
\langle \mu_1,\mu_2\|V_l\|n_2,n_1\rangle
\langle \nu,n_1\|V_l\|\mu_1,\eps\rangle}
{(2l+1)^2(2\ell +1)(E+\eps_{n_2}-\eps_{\nu}-\eps_{\mu_2})
(\eps_{n_1}+\eps_{n_2}-\eps_{\mu_1}-\eps_{\mu_2})},
\end{equation}
%Figure \ref{fig:Sig_3}(d):
% this is a 3rd order exchange self-energy diagram with simultaneous double excitation.
\begin{equation}\label{firs4}
-\sum_{\stackrel{\nu,\mu_1,\mu_2>F}{n_1,n_2\leq F}}\sum_l
\frac{2\langle \eps',n_2\|V_l\|\mu_1,\nu\rangle
\langle \mu_1,\mu_2\|V^{(l)}\|n_1,n_2\rangle
\langle \nu ,n_1\|V_l\|\mu_2,\eps\rangle}
{(2l+1)(2\ell +1)(E+\eps_{n_2}-\eps_{\nu}-\eps_{\mu_1})
(\eps_{n_1}+\eps_{n_2}-\eps_{\mu_1}-\eps_{\mu_2})},
\end{equation}
%Figure \ref{fig:Sig_3}(e):
% this is the 3rd order self-energy diagram with positron-hole repulsion.
\begin{eqnarray}\label{firs1}
&&\!\!\!\!\sum_{\stackrel{\nu_1,\nu_2,\mu >F}{n_1,n_2\leq F}}\sum_{l,l',l''}
(-1)^{l+\ell +l_{n_1}}
\frac{2\langle \eps',n_2\|V_{l''}\|\mu,\nu_2\rangle
\langle \nu_2,n_1\|V_{l'}\|n_2,\nu_1\rangle
\langle \nu_1,\mu\|V_l\|n_1,\eps\rangle}
{(2\ell +1)(E+\eps_{n_2}-\eps_{\nu_2}-\eps_{\mu})
(E+\eps_{n_1}-\eps_{\nu_1}-\eps_{\mu})}
%&\times&
\sj{l_{\nu_1}}{l'}{l_{\nu_2}}{l''}{\ell}{l}
\sj{l}{l'}{l''}{l_{n_2}}{l_{\mu}}{l_{n_1}}
\end{eqnarray}
where $\ell $ is the orbital angular momentum of the incident positron, ``$>\!F$'' indicates summation over the excited electron states (i.e., those above the Fermi level), and ``$\leq \!F$'' indicates summation over the hole states
(i.e., those at or below the Fermi level).

Similarly, the algebraic expressions for the $\Zeff$ diagrams in \fig{fig:diagz}\,(a)--(g), are, respectively:
% this is the 3rd order $\Z$ diagram with electron-hole attraction.
\begin{equation}\label{firsz}
2\sum_{\stackrel{\nu,\mu_1,\mu_2>F}{n_1,n_2\leq F}}\sum_l
\frac{2\langle \eps ,n_2\|\delta_l\|\mu_2,\nu\rangle
\langle \mu_2,n_1\|V^{(l)}\|\mu_1,n_2\rangle
\langle \nu,\mu_1\|V_l\|n_1,\eps\rangle}
{(2l+1)(2\ell +1)(E+\eps_{n_2}-\eps_{\nu}-\eps_{\mu_2})
(E+\eps_{n_1}-\eps_{\nu}-\eps_{\mu_1})},
\end{equation}
%Figure \ref{fig:diagz}(b):
% this is the 3rd order $\Z$ diagram with positron-hole repulsion.
\begin{eqnarray}\label{firsz1}
&&\!\!\!\!\!\!-2\sum_{\stackrel{\nu_1,\nu_2,\mu >F}{n_1,n_2\leq F}}\sum_{l,l',l''}
(-1)^{l+\ell +l_{n_1}}
\frac{2\langle \eps ,n_2\|\delta_{l''}\|\mu,\nu_2\rangle
\langle \nu_2,n_1\|V_{l'}\|n_2,\nu_1\rangle
\langle \nu_1,\mu\|V_l\|n_1,\eps\rangle}
{(2\ell +1)(E+\eps_{n_2}-\eps_{\nu_2}-\eps_{\mu})
(E+\eps_{n_1}-\eps_{\nu_1}-\eps_{\mu})}%\nonumber\\
%&\times&
\sj{l_{\nu_1}}{l'}{l_{\nu_2}}{l''}{\ell}{l}
\sj{l}{l'}{l''}{l_{n_2}}{l_{\mu}}{l_{n_1}}
\end{eqnarray}
%Figure \ref{fig:diagz}(c):
% this is a 3rd order $\Z$ diagram.
\begin{equation}\label{firsz3}
-2\sum_{\stackrel{\nu,\mu_1,\mu_2>F}{n_1,n_2\leq F}}\sum_l
\frac{4\langle \eps ,n_2\|\delta_l\|\mu_2,\nu\rangle
\langle n_1,\mu_2\|V_l\|n_2,\mu_1\rangle
\langle \nu,\mu_1\|V_l\|n_1,\eps\rangle}
{(2l+1)^2(2\ell +1)(E+\eps_{n_2}-\eps_{\nu}-\eps_{\mu_2})
(E+\eps_{n_1}-\eps_{\nu}-\eps_{\mu_1})},
\end{equation}
%Figure \ref{fig:diagz}(d):
% this is the 3rd order $\Z$ diagram with simultaneous double excitation.
\begin{equation}\label{firsz2}
-2\sum_{\stackrel{\nu,\mu_1,\mu_2>F}{n_1,n_2\leq F}}\sum_l
\frac{4\langle \eps ,n_2\|\delta_l\|\mu_2,\nu\rangle
\langle \mu_1,\mu_2\|V_l\|n_2,n_1\rangle
\langle \nu,n_1\|V_l\|\mu_1,\eps\rangle}
{(2l+1)^2(2\ell +1)(E+\eps_{n_2}-\eps_{\nu}-\eps_{\mu_2})
(\eps_{n_1}+\eps_{n_2}-\eps_{\mu_1}-\eps_{\mu_2})},
\end{equation}
%Figure \ref{fig:diagz}(e):
% this is the 3rd order $\Z$ diagram with simultaneous double excitation.
\begin{equation}\label{firszz2}
-2\sum_{\stackrel{\nu,\mu_1,\mu_2>F}{n_1,n_2\leq F}}\sum_l
\frac{4\langle \eps ,n_2\|V_l\|\mu_2,\nu\rangle
\langle \mu_1,\mu_2\|V_l\|n_2,n_1\rangle
\langle \nu,n_1\|\delta_l\|\mu_1,\eps\rangle}
{(2l+1)^2(2\ell +1)(E+\eps_{n_2}-\eps_{\nu}-\eps_{\mu_2})
(\eps_{n_1}+\eps_{n_2}-\eps_{\mu_1}-\eps_{\mu_2})},
\end{equation}
%Figure \ref{fig:diagz}(f):
% this is a 3rd order exchange $\Z$ diagram with simultaneous double excitation.
\begin{equation}\label{firsz5}
2\sum_{\stackrel{\nu,\mu_1,\mu_2>F}{n_1,n_2\leq F}}\sum_l
\frac{2\langle \eps ,n_2\|V_l\|\mu_1,\nu\rangle
\langle \mu_1,\mu_2\|V^{(l)}\|n_1,n_2\rangle
\langle \nu,n_1\|\delta_l\|\mu_2,\eps\rangle}
{(2l+1)(2\ell +1)(E+\eps_{n_2}-\eps_{\nu}-\eps_{\mu_1})
(\eps_{n_1}+\eps_{n_2}-\eps_{\mu_1}-\eps_{\mu_2})},
\end{equation}
%Figure \ref{fig:diagz}(g):
% this is a 3rd order exchange $\Z$ diagram with simultaneous double excitation.
\begin{equation}\label{firsz4}
2\sum_{\stackrel{\nu,\mu_1,\mu_2>F}{n_1,n_2\leq F}}\sum_l
\frac{2\langle \eps ,n_2\|\delta_l\|\mu_1,\nu\rangle
\langle \mu_1,\mu_2\|V^{(l)}\|n_1,n_2\rangle
\langle \nu,n_1\|V_l\|\mu_2,\eps\rangle}
{(2l+1)(2\ell +1)(E+\eps_{n_2}-\eps_{\nu}-\eps_{\mu_1})
(\eps_{n_1}+\eps_{n_2}-\eps_{\mu_1}-\eps_{\mu_2})}.
\end{equation}
%\end{widetext}
%\newpage

%**************************************************************************** 

\section{Tabulated numerical results}\label{app:res}
In order to facilitate comparison of the present calculations with future
experimental and theoretical data we tabulate here the $s$-, $p$- and $d$-wave phase 
shifts, elastic scattering cross sections and $s$-, $p$-, $d$-wave $\Zeff$ 
as a function of the incident positron momentum $k$ for the noble gas atoms.

\begin{table}[htb!]
\begin{ruledtabular}
\caption{Scattering phase shift $\delta_{\ell}$ (in radians), elastic scattering cross section (10$^{-16}$ cm$^2$) and $\Zeff^{(\ell)}$ for $\ell$-wave positrons annihilating on helium. Numbers in brackets denote powers of ten.}
\label{app1a}
\begin{tabular}{cccccccc}
$k$ &\multicolumn{3}{c}{Scattering phase shift}	
&Cross Section & \multicolumn{3}{c}{Annihilation rate} \\
(a.u.) &$\delta _0$ & $\delta _1$ & $\delta _2$ &  (10$^{-16}$ cm$^2$)  	
& $\Zeff^{(s)}$ & $\Zeff^{(p)}$ & $\Zeff^{(d)}$\\
\hline
    0.02  &  8.104[-3]  & 1.076[-4]  & 1.551[-5] & 5.781[-1] & 3.804[0]    & 1.297[-3]   & 1.490[-7]   \\
    0.04  &  1.514[-2]  & 4.176[-4]  & 6.444[-5] & 5.053[-1] & 3.788[0]    & 5.187[-3]   & 2.381[-6]   \\
    0.06  &  2.109[-2]  & 9.067[-4]  & 1.432[-4] & 4.372[-1] & 3.751[0]    & 1.167[-2]   & 1.203[-5]   \\
    0.08  &  2.606[-2]  & 1.568[-3]  & 2.483[-4] & 3.776[-1] & 3.709[0]    & 2.072[-2]   & 3.794[-5]   \\
    0.10  &  3.005[-2]  & 2.403[-3]  & 3.788[-4] & 3.241[-1] & 3.660[0]    & 3.233[-2]   & 9.235[-5]   \\
    0.20  &  3.716[-2]  & 8.784[-3]  & 1.500[-3] & 1.430[-1] & 3.375[0]    & 1.272[-1]   & 1.439[-3]   \\
    0.30  &  2.760[-2]  & 1.778[-2]  & 3.396[-3] & 6.972[-2] & 3.079[0]    & 2.764[-1]   & 6.976[-3]   \\
    0.40  &  6.907[-3]  & 2.795[-2]  & 6.093[-3] & 5.777[-2] & 2.808[0]    & 4.659[-1]   & 2.077[-2]   \\
    0.50  & -2.074[-2]  & 3.795[-2]  & 9.585[-3] & 7.501[-2] & 2.569[0]    & 6.779[-1]   & 4.714[-2]   \\
    0.60  & -5.235[-2]  & 4.667[-2]  & 1.380[-2] & 1.023[-1] & 2.363[0]    & 8.949[-1]   & 8.934[-2]   \\
    0.70  & -8.564[-2]  & 5.340[-2]  & 1.868[-2] & 1.298[-1] & 2.186[0]    & 1.102[0]       & 1.512[-1]   \\
    0.80  & -1.192[-1]  & 5.808[-2]  & 2.402[-2] & 1.535[-1] & 2.032[0]    & 1.297[0]       & 2.302[-1]   \\
    0.90  & -1.519[-1]  & 6.043[-2]  & 2.972[-2] & 1.717[-1] & 1.902[0]    & 1.465[0]       & 3.366[-1]   \\
    1.00  & -1.830[-1]  & 6.113[-2]  & 3.559[-2] & 1.850[-1] & 1.784[0]    & 1.625[0]       & 4.500[-1]
\end{tabular}
\end{ruledtabular}
\end{table}

\begin{table}[htb!]
\begin{ruledtabular}
\caption{Scattering phase shift $\delta_{\ell}$ (in radians), elastic scattering cross section and $\Zeff^{(\ell)}$ for $\ell$-wave positrons on neon. Numbers in brackets denote powers of ten.}
\label{app1b}
\begin{tabular}{ccccccccccc}
$k$ &\multicolumn{3}{c}{Scattering phase shift}	
&Cross Section
&\multicolumn{2}{c}{$\Zeff^{(s)}$}	
&\multicolumn{2}{c}{$\Zeff^{(p)}$}	
&\multicolumn{2}{c}{$\Zeff^{(d)}$}\\
(a.u.) &$\delta _0$ & $\delta _1$ & $\delta _2$	&  (10$^{-16}$ cm$^2$) 	
&$n$\footnote{Total for annihilation on valence $n$ shell ($2s+2p$ subshell total).} 
&$(n-1)$\footnote{Total for annihilation on core $(n-1)$ shell ($1s$ subshell), from Ref.~\cite{DGGinnershells}.} 	
&$n$\footnotemark[1]&$(n-1)$\footnotemark[2]&$n$\footnotemark[1]&$(n-1)$\footnotemark[2]\\
\hline
 0.02    &  8.201[-3]	& 2.081[-4]   & 3.099[-5] 	&5.929[-1]    & 5.590[0]    & 1.612[-2]	& 3.380[-3]	& 1.671[-6]	    & 4.275[-7]       & 7.004[-12]	  \\
 0.04    &  1.443[-2]	& 7.891[-4]   & 1.254[-4] 	&4.622[-1]    & 5.549[0]    & 1.602[-2]	& 1.353[-2]	& 4.674[-6]	    & 6.834[-6]       & 1.121[-10]	  \\
 0.06    &  1.873[-2]	& 1.691[-3]   & 2.759[-4] 	&3.517[-1]	   & 5.465[0]    & 1.579[-2]	& 3.043[-2]	& 1.053[-5]	    & 3.455[-5]       & 5.681[-10]	  \\
 0.08    &  2.133[-2]	& 2.902[-3]   & 4.735[-4] 	&2.648[-1]	   & 5.371[0]    & 1.555[-2]	& 5.408[-2]	& 1.876[-5]	    & 1.090[-4]       & 1.797[-9]	  \\
 0.10    &  2.229[-2]	& 4.421[-3]   & 7.147[-4] 	&1.966[-1]	   & 5.265[0]    & 1.523[-2]	& 8.441[-2]	& 2.936[-5]	    & 2.653[-4]       & 4.392[-9]	  \\
 0.16\footnote{Values given to allow the accurate reproduction of the minimum in the elastic scattering cross-section.} & 1.652[-2]     & 1.067[-2]   & 1.773[-3]	&8.725[-2]   &	--	       &	--	&	--		&	--		    &		--	   &	--		  \\	
 0.18\footnotemark[3]    &  1.211[-2]    & 1.320[-2]   & 2.253[-3] 	&7.628[-2]   &	--	       &	--	&	--		&	--		    &		--	   &	--		  \\	
 0.20    &  6.652[-3]	& 1.592[-2]   & 2.792[-3] 	& 7.522[-2]  & 4.695[0]    & 1.387[-2]	& 3.314[-1]	& 1.178[-4]	    & 4.141[-3]       & 7.062[-8]	  \\
 0.22\footnotemark[3]     & 2.191[-4]    & 1.880[-2]   & 3.384[-3] 	& 8.245[-2]  &	--	       &	--	&	--		&	--		    &		--	   &	--		  \\	 
 0.24\footnotemark[3]     &-7.089[-3]    & 2.181[-2]   & 4.035[-3] 	& 9.668[-2]  &	--	       &	--	&	--		&	--		    &		--	   &	--		  \\
 0.30    & -3.347[-2]	& 3.130[-2]   & 6.383[-3] 	&1.689[-1]	   & 4.169[0]    & 1.268[-2]	& 7.122[-1]	& 2.626[-4]	    & 2.005[-2]       & 3.577[-7]	  \\
 0.40    & -8.843[-2]	& 4.692[-2]   & 1.155[-2] 	&3.354[-1]	   & 3.729[0]    & 1.180[-2]	& 1.174[0] 	& 4.549[-4]	    & 5.955[-2]       & 1.131[-6]	  \\
 0.50    & -1.520[-1]	& 5.943[-2]   & 1.824[-2] 	&5.014[-1]	   & 3.368[0]    & 1.120[-2]	& 1.656[0] 	& 6.817[-4]	    & 1.344[-1]       & 2.779[-6]	  \\
 0.60    & -2.200[-1]	& 6.655[-2]   & 2.618[-2] 	&6.378[-1]	   & 3.074[0]    & 1.084[-2]	& 2.107[0] 	& 9.297[-4]	    & 2.529[-1]       & 5.661[-6]	  \\
 0.70    & -2.897[-1]	& 6.700[-2]   & 3.503[-2] 	&7.390[-1]	   & 2.834[0]    & 1.066[-2]	& 2.492[0] 	& 1.191[-3]	    & 4.248[-1]       & 1.020[-5]	  \\
 0.80    & -3.597[-1]	& 6.093[-2]   & 4.406[-2] 	&8.119[-1]	   & 2.632[0]    & 1.065[-2]	& 2.810[0] 	& 1.461[-3]	    & 6.338[-1]       & 1.756[-5]	  \\
 0.90    & -4.283[-1]	& 4.851[-2]   & 5.289[-2] 	&8.612[-1]	   & 2.469[0]    & 1.079[-2]	& 3.034[0] 	& 1.733[-3]	    & 9.162[-1]       & 2.803[-5]	  \\
 1.00    & -4.947[-1]	& 3.134[-2]   & 6.080[-2] 	&8.938[-1]	   & 2.326[0]    & 1.105[-2]	& 3.236[0] 	& 2.011[-3]	    & 1.187[0]	      & 4.045[-5]	  \\
\end{tabular}
\end{ruledtabular}
\end{table}

\begin{table}[htb!]
\begin{ruledtabular}
\caption{Scattering phase shift $\delta_{\ell}$ (in radians), elastic scattering cross section and $\Zeff^{(\ell)}$ for $\ell$-wave positrons on argon. Numbers in brackets denote powers of ten.}
\label{app1c}
\begin{tabular}{ccccccccccc}
$k$ &\multicolumn{3}{c}{Scattering phase shift}	
&Cross Section
&\multicolumn{2}{c}{$\Zeff^{(s)}$}	
&\multicolumn{2}{c}{$\Zeff^{(p)}$}	
&\multicolumn{2}{c}{$\Zeff^{(d)}$}\\
(a.u.) &$\delta _0$ & $\delta _1$ & $\delta _2$	&  (10$^{-16}$ cm$^2$) 	
&$n$\footnote{Total for annihilation on valence $n$ shell ($3s+3p$ subshell total).} 
&$(n-1)$\footnote{Total for annihilation on core $(n-1)$ shell ($2s+2p$ subshell total), from Ref.~\cite{DGGinnershells}.} 	
&$n$\footnotemark[1]&$(n-1)$\footnotemark[2]&$n$\footnotemark[1]&$(n-1)$\footnotemark[2]\\
\hline
 0.02   &    8.178[-2]	  &  8.777[-4]   &  1.275[-4]&  5.873[1]    &   2.865[1]  & 1.653[-1]	 &  1.992[-2]	      & 3.397[-5]     & 4.288[-6]  & 9.621[-10]  \\
 0.04   &    1.502[-1]	  &  3.453[-3]   &  5.099[-4]&  4.933[1]    &   2.698[1]  & 1.558[-1]	 &  8.007[-2]	      & 1.368[-4]     & 6.858[-5]  & 1.542[-8]   \\
 0.06   &    2.011[-1]	  &  7.673[-3]   &  1.132[-3]&  3.918[1]   &   2.432[1]  & 1.407[-1]	 &  1.812[-1]	      & 3.104[-4]     & 3.469[-4]  & 7.825[-8]   \\
 0.08   &    2.374[-1]	  &  1.356[-2]   &  1.980[-3]&  3.073[1]   &   2.176[1]  & 1.261[-1]	 &  3.239[-1]	      & 5.569[-4]     & 1.094[-3]  & 2.481[-7]   \\
 0.10   &    2.604[-1]	  &  2.112[-2]   &  3.060[-3]&  2.382[1]   &   1.935[1]  & 1.126[-1]	 &  5.083[-1]	      & 8.782[-4]     & 2.665[-3]  & 6.077[-7]   \\
 0.20   &    2.539[-1]	  &  7.964[-2]   &  1.291[-2]& 7.312[0]	&   1.146[1]  & 6.873[-2]     &  1.998[0]  	   & 3.589[-3]     & 4.149[-2]  & 9.918[-6]   \\
 0.30   &    1.534[-1]	  &  1.548[-1]   &  3.096[-2]& 3.928[0]	&   7.754[0]     & 4.900[-2]     &  3.995[0]  	   & 7.632[-3]     & 1.988[-1]  & 5.104[-5]  \\
 0.40   &    2.149[-2]	  &  2.172[-1]   &  5.835[-2]& 3.518[0]	&   5.785[0]     & 3.929[-2]     &  5.670[0]  	   & 1.176[-2]     & 5.753[-1]  & 1.625[-4]  \\
 0.50   &   -1.195[-1]	  &  2.484[-1]   &  9.424[-2]& 3.484[0]	&   4.608[0]     & 3.424[-2]     &  6.595[0]  	   & 1.510[-2]     & 1.240[0]	& 4.051[-4]    \\
 0.60   &   -2.603[-1]	  &  2.464[-1]   &  1.352[-1]& 3.437[0]	&   3.834[0]     & 3.165[-2]     &  6.887[0]  	   & 1.770[-2]     & 2.183[0]	& 7.911[-4]    \\
 0.70   &   -3.966[-1]	  &  2.181[-1]   &  1.773[-1]& 3.410[0]	&   3.306[0]     & 3.066[-2]     &  6.818[0]  	   & 1.976[-2]     & 3.332[0]	& 1.588[-3]    \\
 0.80   &   -5.273[-1]	  &  1.732[-1]   &  2.175[-1]& 3.440[0]	&   2.916[0]     & 3.068[-2]     &  6.601[0]  	   & 2.217[-2]     & 4.673[0]	& 2.268[-3]
\end{tabular}
\end{ruledtabular}
\end{table}

\begin{table}[htb!]
\begin{ruledtabular}
\caption{Scattering phase shift $\delta_{\ell}$, elastic scattering cross section and $\Zeff^{(\ell)}$ for $\ell$-wave positrons on krypton. Numbers in brackets denote powers of ten.}
\label{app1d}
\begin{tabular}{ccccccccccc}
$k$ &\multicolumn{3}{c}{Scattering phase shift}	
&Cross Section
&\multicolumn{2}{c}{$\Zeff^{(s)}$}	
&\multicolumn{2}{c}{$\Zeff^{(p)}$}	
&$\Zeff^{(d)}$\\
(a.u.) &$\delta _0$ & $\delta _1$ & $\delta _2$	&  (10$^{-16}$ cm$^2$) 	
&$n$\footnote{Total for annihilation on valence $n$ shell ($4s+4p$ subshell total).} 
&$(n-1)$\footnote{Total for annihilation on core $(n-1)$ shell ($3s+3p+3d$ subshell total), from Ref.~\cite{DGGinnershells}.} 	
&$n$\footnotemark[1]&$(n-1)$\footnotemark[2]&$n$\footnotemark[1]\\
 0.02  &   1.812[-1]  &   1.378[-3]  &  1.990[-4]& 2.859[2] &  8.419[1]    & 1.362[0]	 &  3.939[-2]	 & 2.438[-4]  &  9.621[-6]	 \\
 0.04  &   3.228[-1]  &   5.454[-3]  &  7.932[-4]& 2.215[2] &  7.112[1]    & 1.150[0]	 &  1.589[-1]	 & 9.855[-4]  &  1.539[-4]     \\
 0.06  &   4.114[-1]  &   1.223[-2]  &  1.757[-3]& 1.568[2] &  5.573[1]    & 9.007[-1]   &  3.615[-1]	 & 2.249[-3]  &  7.788[-4]	 \\
 0.08  &   4.639[-1]  &   2.184[-2]  &  3.065[-3]& 1.109[2] &  4.394[1]    & 7.108[-1]   &  6.500[-1]	 & 4.061[-3]  &  2.458[-3]	 \\
 0.10  &   4.885[-1]  &   3.436[-2]  &  4.731[-3]& 7.880[1] &  3.509[1]    & 5.691[-1]   &  1.026[0]	 & 6.444[-3]  &  5.990[-3]	\\
 0.20  &   4.269[-1]  &   1.338[-1]  &  2.038[-2]& 2.000[1] &  1.526[1]    & 2.556[-1]   &  4.045[0]	 & 2.657[-2]  &  9.364[-2]    \\
 0.30  &   2.631[-1]  &   2.560[-1]  &  5.050[-2]& 1.075[1] &  9.063[0]	   & 1.610[-1]   &  7.527[0]	 & 5.293[-2]  &  4.501[-1]    \\
 0.40  &   7.878[-2]  &   3.399[-1]  &  9.779[-2]& 8.678[0] &  6.318[0]    & 1.219[-1]   &  9.384[0]     & 7.220[-2]  &  1.294[0]	  \\
 0.50  &  -1.053[-1]  &   3.625[-1]  &  1.596[-1]& 7.492[0] &  4.831[0]	   & 1.031[-1]   &  9.584[0]     & 8.203[-2]  &  2.718[0]    \\
 0.60  &  -2.826[-1]  &   3.352[-1]  &  2.273[-1]& 6.774[0] &  3.915[0]	   & 9.391[-2]   &  9.020[0]     & 8.751[-2]  &  4.548[0]	\\
 0.70  &  -4.493[-1]  &   2.777[-1]  &  2.933[-1]& 6.463[0] &  3.309[0]	   & 9.096[-2]   &  8.316[0]     & 9.098[-2]  &  6.627[0]	\\
\end{tabular}
\end{ruledtabular}
\end{table}

\begin{table}[htb!]
\begin{ruledtabular}
\caption{Scattering phase shift $\delta_{\ell}$ (in radians), elastic scattering cross section and $\Zeff^{(\ell)}$ for 
$\ell$-wave positrons on xenon. Numbers in brackets denote powers of ten.}
\label{app1e}
\begin{tabular}{cccccccccc}
$k$ &\multicolumn{3}{c}{Scattering phase shift}	
&Cross Section
&\multicolumn{2}{c}{$\Zeff^{(s)}$}	
&\multicolumn{2}{c}{$\Zeff^{(p)}$}	
&$\Zeff^{(d)}$\\
(a.u.) &$\delta _0$ & $\delta _1$ & $\delta _2$	&  (10$^{-16}$ cm$^2$) 	
&$n$\footnote{Total for annihilation on valence $n$ shell ($5s+5p$ subshell total).} 
&$(n-1)$\footnote{Total for annihilation on core $(n-1)$ shell ($4s+4p+4d$ subshell total), from Ref.~\cite{DGGinnershells}.} 	
&$n$\footnotemark[1]&$(n-1)$\footnotemark[2]&$n$\footnotemark[1]\\
  0.02  &   9.849[-1]   &  2.303[-3]   &  3.223[-4]&6.108[3] &  1.227[3] & 2.971[1]  &  1.040[-1]    & 1.085[-3]  &  2.716[-5]    \\
  0.04  &   1.151[0]    &  9.418[-3]   &  1.285[-3]&1.835[3] &  3.517[2] & 8.289[0]  &  4.235[-1]    & 4.428[-3]  &  4.350[-4]    \\
  0.06  &   1.130[0]    &  2.178[-2]   &  2.856[-3]&8.010[2] &  1.610[2] & 3.767[0]  &  9.763[-1]    & 1.025[-2]  &  2.205[-3]    \\
  0.08  &   1.085[0]    &  4.001[-2]   &  5.015[-3]&4.327[2] &  9.391[1] & 2.195[0]  &  1.782[0]     & 1.880[-2]  &  6.975[-3]    \\
  0.10  &   1.027[0]    &  6.450[-2]   &  7.809[-3]&2.622[2] &  6.239[1] & 1.463[0]  &  2.854[0]     & 3.032[-2]  &  1.704[-2]    \\
  0.20  &   7.199[-1]   &  2.666[-1]   &  3.572[-2]&5.723[1] &  1.888[1] & 4.608[-1] &  1.122[1]     & 1.256[-1]  &  2.713[-1]    \\
  0.30  &   4.256[-1]   &  4.784[-1]   &  9.425[-2]&3.349[1] &  1.001[1] & 2.623[-1] &  1.675[1]     & 2.023[-1]  &  1.325[0]	  \\
  0.40  &   1.551[-1]   &  5.624[-1]   &  1.904[-1]&2.365[1] &  6.618[0] & 1.912[-1] &  1.584[1]     & 2.110[-1]  &  3.769[0]	  \\
  0.50  &  -9.220[-2]   &  5.354[-1]   &  3.118[-1]&1.839[1] &  4.906[0] & 1.597[-1] &  1.328[1]     & 1.989[-1]  &  7.384[0]	  \\
  0.60  &  -3.192[-1]   &  4.503[-1]   &  4.298[-1]&1.595[1] &  3.896[0] & 1.453[-1] &  1.108[1]     & 1.896[-1]  &  1.099[1]	  \\
\hline\hline
\end{tabular}
\end{ruledtabular}
\end{table}
%\end{widetext}


\begin{thebibliography}{10}
\bibitem{phase} See also Ref.~\cite{Amusia2}; in this paper the self-energy operator $\Sigma _\eps$ is expressed in Rydberg and the coefficient in \eqn{eq:tandel} is $-\pi $.
\bibitem{comm:Zeff}The many-body theory $\Zeff$ in \fig{fig:xezeff_fit} contains only the contributions of the positron $s$-, $p$-, and $d$-waves.
At low momenta $k$ the contribution of each next partial wave is suppressed by the factor $\propto k^2$. By comparing the $p$- and $d$-wave $\Zeff$ values, we estimate that the contribution of all higher partial waves to the total $\Zeff$ should not exceed $\sim 4$ at $k=0.6$~a.u., and is smaller at lower energies.
\end{thebibliography}
\end{document}